\documentclass[final,5p,times,twocolumn]{elsarticle}

\usepackage{graphicx}
\usepackage{url}
\usepackage{array}
\usepackage{color}
\usepackage{amsmath, amssymb}
\usepackage[dvipsnames]{xcolor}

\newcounter{bla}

\journal{Computer Physics Communications}

\begin{document}

\newcommand{\vectornorm}[1]{\left|#1\right|}
\definecolor{bg}{rgb}{0.95,0.95,0.95}

\begin{frontmatter}

  \title{Afivo: a framework for quadtree/octree AMR with shared-memory
    parallelization and geometric multigrid methods}

  \author[a,b]{Jannis Teunissen\corref{jannis}}
  \author[b,c]{Ute Ebert}

  \cortext[jannis] {Corresponding author.\\\textit{E-mail address:}
    jannis@teunissen.net}

  \address[a]{Centre for Mathematical Plasma-Astrophysics, KU Leuven,
    Celestijnenlaan 200B, 3001 Leuven, Belgium}
  \address[b]{Centrum Wiskunde \& Informatica, PO Box 94079, 1090 GB Amsterdam,
    The Netherlands}
  \address[c]{Department of Applied Physics, Eindhoven University of Technology,
    PO Box 513, 5600 MB Eindhoven, The Netherlands}

  \begin{abstract}
    Afivo is a framework for simulations with adaptive mesh refinement (AMR) on
    quadtree (2D) and octree (3D) grids. The framework comes with a geometric
    multigrid solver, shared-memory (OpenMP) parallelism and it supports output
    in Silo and VTK file formats. Afivo can be used to efficiently simulate AMR
    problems with up to about $10^{8}$ unknowns on desktops, workstations or single
    compute nodes. For larger problems, existing distributed-memory frameworks
    are better suited. The framework has no built-in functionality for specific
    physics applications, so users have to implement their own numerical
    methods. The included multigrid solver can be used to efficiently solve
    elliptic partial differential equations such as Poisson's equation. Afivo's
    design was kept simple, which in combination with the shared-memory
    parallelism facilitates modification and experimentation with AMR
    algorithms. The framework was already used to perform 3D simulations of
    streamer discharges, which required tens of millions of cells.
  \end{abstract}


\end{frontmatter}


{\bf PROGRAM SUMMARY}

\begin{small}
  \noindent
  {\em Manuscript Title:} Afivo: a framework for quadtree/octree AMR with
  shared-memory parallelization and geometric multigrid methods\\
  {\em Authors:} Jannis Teunissen and Ute Ebert\\
  {\em Program Title:} Afivo\\
  {\em Journal Reference:}                                      \\
  {\em Catalogue identifier:}                                   \\
  {\em Licensing provisions:} GPLv3\\
  {\em Programming language:} Fortran 2011 \\
  {\em Operating system:} Unix-like systems with a Fortran \& C compiler\\
  {\em RAM:} 1 MB up to tens of GB\\
  {\em Number of processors used:} OpenMP is supported 
  {\em Classification:}
  4.12 Other numerical methods, 6.5 Software including Parallel Algorithms, 19.5
  Discharges\\
  {\em External routines/libraries:} Silo (LLNL)\\
  {\em Nature of problem:}\\
  Performing multiscale simulations, especially those requiring a fast elliptic solver. \\
  {\em Solution method:}\\
  Provide a framework for parallel simulations on adaptively refined
  quadtree/octree grids, including a geometric multigrid solver.
  \\
  {\em Unusual features:}\\
  The framework uses shared-memory parallelism (OpenMP) instead of MPI.
  \\
  {\em Running time:}\\
  Linear in the number of unknowns, with a small constant. \\
\end{small}

\section{Introduction}
\label{sec:intro}

Many systems have a \emph{multiscale} nature, meaning that physical structures
occur at different spatial and temporal scales. These structures can appear at
different locations and move in space. Numerical simulations of such systems can
be speed up with adaptive mesh refinement (AMR), especially if a fine mesh is
required in only a small part of the domain. Here we present Afivo (Adaptive
Finite Volume Octree), a framework for simulations with AMR on structured grids.
Some of the key characteristics of Afivo are
\begin{itemize}
  \item Adaptively refined quadtree (2D) and octree (3D) grids
  \item OpenMP parallelization
  \item A geometric multigrid solver
  \item Output in Silo and VTK file formats
  \item Source code in Fortran 2011 with \texttt{GNU GPLv3} license
\end{itemize}

An overview of Afivo's functionality and potential applications is given below,
together with a brief discussion of our motivation for developing the framework.
An overview of the design, data structures and methods is given in
section~\ref{sec:data-procedures}. An important part is the geometric multigrid
solver, which handles refinement boundaries in a consistent way. The
implementation of this solver is described in section \ref{sec:multigrid}.
Finally, some examples are presented in section \ref{sec:examples}.

\subsection{Overview and applications}
\label{sec:afivo-basics}

As a generic simulation framework, Afivo comes without solvers for specific
physics problems. A user thus has to implement the required numerical methods as
well as a suitable refinement criterion, see section \ref{sec:ref-procedure}. We
think Afivo could be used when one wants to investigate numerical
discretizations or AMR algorithms, or when no existing simulation software is
available for the problem at hand. To demonstrate some of the framework's
possibilities, several examples are included in the \texttt{examples} directory
of the source code:
\begin{itemize}
  \item Examples showing e.g., how to define the computational domain, perform
  refinement, set boundary conditions and write output.
  \item Solving a scalar advection equation in 2D and 3D using the explicit
  trapezoidal rule and the Koren flux limiter \cite{koren_limiter}.
  \item Solving a time-dependent 2D diffusion/heat equation implicitly using the
  backward Euler method and geometric multigrid routines.
  \item Solving a Laplace/Poisson equation on a Cartesian grid (2D, 3D) or in
  cylindrical $(r,z)$ coordinates with geometric multigrid.
  \item Simulating a destabilizing ionization wave in 2D, see section
  \ref{sec:example-discharge}.
  \item Mapping particles to densities on a mesh, and interpolating
  mesh variables to particles.
\end{itemize}

Dirichlet, Neumann and periodic boundary conditions are supported, but other
types of boundary conditions can easily be added. For Dirichlet and Neumann
conditions, the user has to provide a routine that specifies the value of the
solution/derivative at the boundary. Boundary conditions are implemented through
ghost cells, so that numerical methods do not have to be modified near the
boundary of a grid block, see section \ref{sec:basics-grid}. For the same
reason, values from neighboring blocks are also communicated through ghost
cells. It is the user's responsibility to ensure that ghost cells are up to
date, see section \ref{sec:ghost-cell}.

Afivo is most suited for relatively low order spatial discretizations, e.g.
second or third order. The corresponding numerical operators have a small
stencil, which reduces the communication overhead due to the adaptive grid.
Shared-memory parallelism is employed, which means one can experiment with
different AMR methods without having to deal with load balancing or the
communication between processors. This is for example relevant when comparing
different schemes to fill ghost cells near refinement boundaries. With
shared-memory parallelism, Afivo can still be used for problems with tens of
millions of unknowns as current hardware often provides 16 or more CPU cores
with at least as many gigabytes of RAM.

\subsection{Source code and documentation}
\label{sec:source-docs}

Afivo is written in modern Fortran, using some of the features of Fortran 2011.
The 2D and 3D version of the framework are automatically generated from a set of
common source files, using a preprocessor. For example, the module
\texttt{src/m\_aX\_ghostcell.f90}, which contains methods for filling ghost
cells, is translated to \texttt{m\_a2\_ghostcell.f90} (2D) and
\texttt{m\_a3\_ghostcell.f90} (3D). Most of Afivo's methods and types have a
prefix \texttt{a2\_} in 2D and \texttt{a3\_} in 3D. The source code is
documented using Doxygen, and it comes with a brief user guide. An online
version of this documentation is available through
\url{https://gitlab.com/MD-CWI-NL/afivo}.

\subsection{The grid and refinement}
\label{sec:basics-grid}

\begin{figure*}
  \centering
  \includegraphics[width=0.9\textwidth]{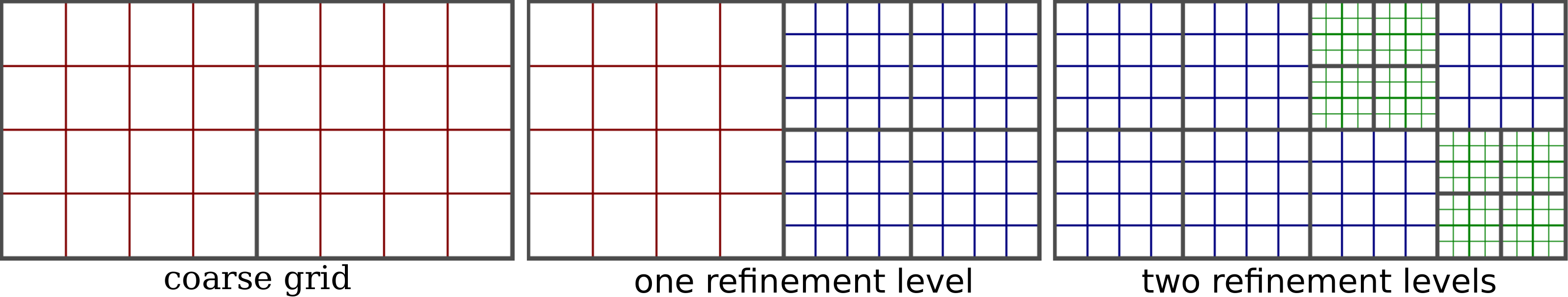}
  \caption{Left: example of a coarse grid consisting of two boxes of
    $4 \times 4$ cells. The middle and right figure show how the boxes can be
    refined, by covering them with four `child' boxes.}
  \label{fig:quadtree-comp-domain}
\end{figure*}

Afivo uses a quadtree/octree grid. For simplicity, the description below is for
quadtrees in 2D, the generalization to octrees in 3D is straightforward. A
quadtree grid in Afivo consists of \emph{boxes} (i.e., blocks) of $N \times N$
cells, with $N$ an even number. A user can for example select to use boxes of
$4 \times 4$ cells. The coarse grid, which defines the computational domain, can
then be constructed from one or more of these boxes, see
figure~\ref{fig:quadtree-comp-domain} and section \ref{sec:coarse-grid}.

Two types of variables are stored: cell-centered variables and face-centered
variables. When initializing Afivo, the user has to specify how many of these
variables are required. For the cell-centered variables, each box has one layer
of ghost cells, as discussed in section~\ref{sec:ghost-cell}. For each
cell-centered variable, users can specify default procedures for filling ghost cells
and to perform interpolation and restriction.

A box in a quadtree grid can be refined by adding four \emph{child} boxes.
These children contain the same number of cells but half the grid spacing, so
that they together have the same area as their parent. Each of the children can
again be refined, and so on, as illustrated in figure
\ref{fig:quadtree-comp-domain}. There can be up to 30 refinement levels in
Afivo. So-called \emph{proper nesting} or \emph{2:1 balance} is ensured, which
means that neighboring boxes differ by at most one refinement level.

Afivo does not come with built-in refinement criteria. Instead, users have to
supply a routine that sets refinement flags for the cells in a box. There are
three possible flags: refine, derefine or keep the current refinement. The
user's routine is then automatically called for all relevant boxes of the grid,
after which boxes are refined and derefined, see section \ref{sec:ref-procedure}
for details. For simplicity, each mesh adaptation can locally change the
refinement level at a location by at most one. After the grid has been modified,
the user gets information on which boxes have been removed and which ones have
been added.

For each refinement level, Afivo stores three lists: one with all the parent
boxes, one with all the leaves (which have no children), and one with both
parents and leaves. To perform computations on the boxes, a user can loop over
the levels and over these lists in a desired order. Because of the shared memory
parallelism, values on the neighbors, parents or children of a box can always be
accessed, see section \ref{sec:data-procedures} for details.

\subsection{Motivation and alternatives}
\label{sec:alternatives-design}

There already exist numerous parallel AMR frameworks that operate on structured
grids, some of which are listed in table~\ref{tab:amr-frameworks}. Some of these
frameworks use block-structured grids\footnote{A reviewer pointed out
  that SAMRAI, BoxLib, and Chombo can also be used with octree grids.}, in which
grid blocks can have different sizes (in terms of number of cells). Any octree
mesh is also a block-structured mesh, whereas the opposite is not true. The
connectivity of an octree mesh is simpler, because each block has the same
number of cells, and blocks are always refined in the same way.


We were interested in AMR frameworks that could be used for simulations of
\emph{streamer discharges} (e.g., \cite{Vitello_1994,Yi_2002,Ebert_2010}). Such
simulations require a fine mesh where the streamers grow, and at every time step
Poisson's equation has to be solved to compute the electric field. In
\cite{Pancheshnyi_2008}, Paramesh was used for streamer simulations, but the
main bottleneck was the Poisson solver. Other streamer models (see e.g.,
\cite{Montijn_2006, Li_hybrid_ii_2012, Luque_2012}) faced the same challenge,
because the non-local nature of Poisson's equation makes an efficient parallel
solution difficult, especially on adaptively refined grids. Geometric multigrid
methods can overcome most of these challenges, as demonstrated
in~\cite{Kolobov_2012}, which adapted its multigrid methods from the Gerris Flow
Solver~\cite{Popinet_2003}. Afivo's multigrid implementation is discussed in
section \ref{sec:multigrid}. Successful applications of Afivo to 3D
streamer simulations can be found in \cite{Nijdam_Teunissen_2016,teunissen_2017_afivo_streamer}.

\begin{table*}
  \centering
  \small
  \begin{tabular}{l|l l l l}
    Name & Application & Language & Parallel & Mesh \\
    \hline
    Boxlib \cite{Zhang_2016} & General & C/F90 & MPI/OpenMP & Block-str. \\
    Chombo \cite{chombo} & General & C++/Fortran & MPI & Block-str. \\
    AMRClaw & Flow & F90/Python & MPI/OpenMP & Block-str. \\
    SAMRAI \cite{Hornung_2006} & General & C++ & MPI & Block-str. \\
    AMROC & Flow & C++ & MPI & Block-str. \\
    Paramesh \cite{Macneice_2000} & General & F90 & MPI & Octree \\
    Dendro \cite{Sampath_2008} & General & C++ & MPI & Octree \\
    Peano \cite{peano_book} & General & C++ & MPI/OpenMP & Octree \\
    Gerris \cite{Popinet_2003} & Flow & C & MPI & Octree \\
    Ramses \cite{Teyssier_2002} & Self gravitation & F90 & MPI & Octree
  \end{tabular}
  \caption{An incomplete list of frameworks for parallel numerical computations
    on adaptively refined but structured numerical grids. For each framework,
    the typical application area, programming language, parallelization method
    and mesh type are listed. This list is largely taken from Donna Calhoun's
    homepage \cite{www_donna_calhoun}.}
  \label{tab:amr-frameworks}
\end{table*}

Several of the framework listed in table \ref{tab:amr-frameworks} include
multigrid solvers, for example Boxlib, Dendro, Gerris and Ramses. Afivo is
different because it is based on shared-memory parallelism and because it is
physics-independent (which e.g., Gerris and Ramses are not). Simulations with
adaptive mesh refinement often require some experimentation, for example to
determine a suitable refinement criterion, to compare multigrid algorithms or to
investigate different discretizations near refinement boundaries. Afivo was
designed to facilitate such experiments, by keeping the implementation
relatively simple:
\begin{itemize}
  \item Only shared-memory parallelism is supported, so that data can be
  accessed directly and no parallel communication or load balancing is required.
  Note that all of the frameworks listed in table~\ref{tab:amr-frameworks} use
  MPI (distributed-memory parallelism).
  \item Quadtree and octree grids are used, which are probably the simplest
  grids that support adaptive refinement.
  \item Only cell-centered and face-centered variables are supported.
  \item The cell-centered variables always have one layer of ghost cells (but
  more can be obtained).
  \item Afivo is application-independent, i.e., it includes no code or
  algorithms for specific applications.
\end{itemize}
Because of these simplifications we expect that Afivo can easily be modified,
thus providing an option in between the `advanced' distributed-memory codes of
table \ref{tab:amr-frameworks} and uniform grid computations.

\section{Afivo data types and procedures}
\label{sec:data-procedures}

The most important data types and procedures used in Afivo are described below.
Not all details about the implementation can be given here; further information
can be found in the code's documentation.

\subsection{The tree, levels and boxes}
\label{sec:tree-lvl-box}

The full quadtree/octree grid is contained in a single Fortran type named
\texttt{a2\_t/a3\_t} in 2D/3D (see the code's documentation for details). All
the boxes are stored in a one-dimensional array, so that each box can be
identified by an integer index.
For each refinement level $l$ up to the maximum level of 30, three lists are
stored:
\begin{itemize}
  \item One with all the boxes at refinement level $l$.
  \item One with the \emph{parents} (boxes that are refined) at level $l$.
  \item One with the \emph{leaves} (boxes that are not refined) at level $l$.
\end{itemize}
This separation is often convenient, because some algorithms operate only on
leaves while others operate on parents or on all boxes. Other information, such
as the highest refinement level, the number of cells in a box, the number of
face and cell-centered variables and the coarse grid spacing is also stored.

\begin{figure}
  \centering
  \includegraphics[width=5cm]{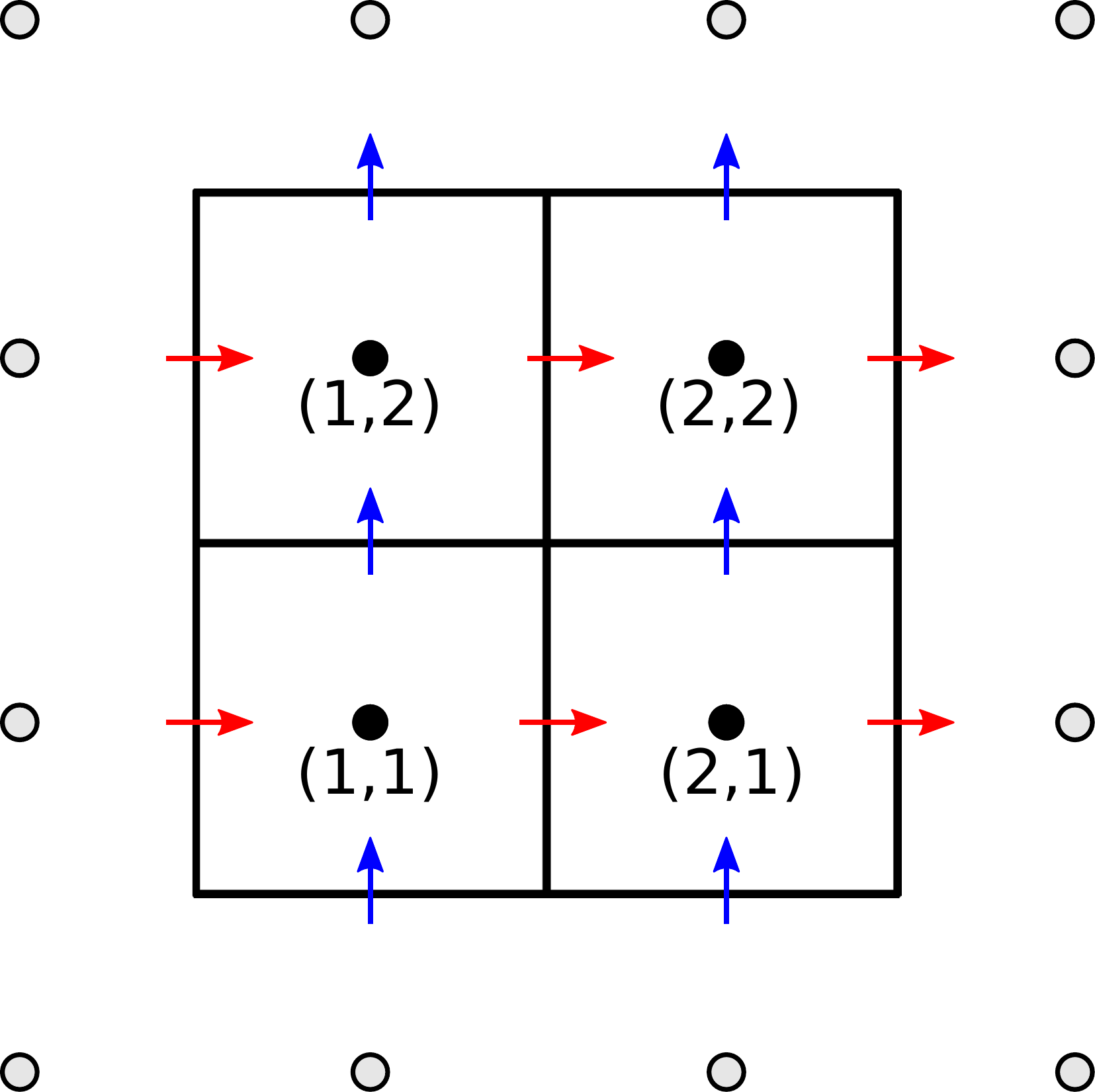}
  \caption{Location and indices of the cell-centered variables (black dots) and the
  face-centered variables in the $x$-direction (red dots) and $y$-direction
  (blue dots) for a box of $2\times 2$ cells. The single layers of ghost cells
  for the cell-centered variables is shown in gray.}
  \label{fig:location-cc-fx}
\end{figure}

When initializing the tree, the user specifies how many cell-centered and
face-centered variables have to be stored. Each box contains one layer of ghost
cells for its cell-centered variables, see figure \ref{fig:location-cc-fx} and
section \ref{sec:ghost-cell}. Furthermore, the indices of the box's parent, its
children and its neighbors (including diagonal ones) are stored. A
special value of zero is used to indicate that a box does not exist, and
negative numbers are used to indicate physical boundaries.

For convenience, boxes also contain information about their refinement level,
their minimum coordinate (e.g., lower left corner in 2D) and their spatial
index. The spatial index of a box defines where the box is located, with $(1,1)$
in 2D or $(1,1,1)$ in 3D being the lowest allowed index. A box with index
$(i,j)$ has neighbors with indices $(i \pm 1, j)$ and $(i, j \pm 1)$, and
children with indices $(2i - 1, 2j - 1)$ up to $(2i, 2j)$.

We remark that the box size $N$ (i.e., it contains $N^D$ cells) should typically
be $8$ or higher, to reduce the overhead of storing neighbors, children, ghost
cells and other information.

\subsection{Defining the computational domain/coarse grid}
\label{sec:coarse-grid}

After initializing the octree data structure, the user can specify a coarse grid
consisting of one or more boxes together with their connectivity. To place two
boxes next to each other, as in the example of figure
\ref{fig:quadtree-comp-domain}, one could place the first one at index $(1,1)$
and the second one at $(2,1)$. If the neighbors of these two boxes are set to
the special value \texttt{af\_no\_box}, their connectivity is automatically
resolved. A periodic boundary in the $x$-direction can be imposed by specifying
that the left neighbor of the first box is box two, and that the right neighbor
of the second box is box one. External boundaries can be indicated by negative
values. Besides rectangular grids, it is also possible to generate e.g.,
L-shaped meshes or O-shaped meshes containing a hole.

\subsection{Mesh refinement}
\label{sec:ref-procedure}

To adapt the mesh, the user has to specify a refinement routine. Given a box,
this routine should specify a refinement flag for each cell: refine, derefine or
keep refinement. Each mesh adaptation changes the mesh by at most one level at a
given location. Boxes are either fully refined (with $2^D$ children) or not
refined, but never partially refined. A number of rules is used to convert the
cell-flags to refinement flags for the boxes:
\begin{itemize}
  \item If any cell in a box is flagged for refinement, the box is refined. If
  neighbors of the box are at a lower refinement level, they are also refined to
  ensure 2:1 balance.
  \item Neighboring boxes within a distance of $N_\mathrm{buf}$ (default: two)
  cells of a cell flagged for refinement will also be refined. This also applies
  to diagonal neighbors.
  \item If all the cells in a box are marked for derefinement, then the box is
  marked for removal, but whether this happens depends on the points below:
  \begin{itemize}
    \item If all the $2^D$ children of a box are flagged for removal, and the box
    itself not for refinement, then the children are removed.
    \item Only leaves can be removed (because the grid changes by at most one
    level at a time).
    \item Boxes cannot be removed if that would violate 2:1 balance.
  \end{itemize}
\end{itemize}

When boxes are added or removed in the refinement procedure, the mesh
connectivity is automatically updated, and the array containing all the boxes is
automatically resized when necessary. The removal of boxes can create holes in
this array, which are automatically filled when their number exceeds a
threshold. The boxes are then also sorted (per level) according to their Morton
index~\cite{Morton_1966}.

If a user has specified routines for prolongation (interpolation) and
restriction of a cell-centered variable, then these operations are automatically
performed when changing the mesh. The built-in prolongation and restriction
routines are described in section \ref{sec:interp-restrict}. After updating the
refinement, information on the added and removed boxes per level is returned, so
a user can also manually set values on new boxes.

\subsection{Ghost cells}
\label{sec:ghost-cell}

The usage of ghost cells has two main advantages: algorithms can operate without
special care for the boundaries, and they can do so in parallel. In Afivo each
box always has one layer of ghost cells around its cell-centered variables. For
numerical operations that depend on the nearest neighbors, such as computing a
second order Laplacian or centered differences, one ghost cell is enough. When
additional ghost values are required, these can directly be accessed due to the
shared-memory parallelism.

Afivo supports a single layer of ghost cells around boxes, including corners
(and edges in 3D). A user has to provide two routines for filling ghost cells on
the sides of boxes, one for physical boundaries and one for refinement
boundaries. A couple of such routines are also included with the framework. For
each box, ghost cells are first filled on the sides, then on the edges (only
present in 3D) and finally on the corners. For each side of a box, there are
three options:
\begin{itemize}
  \item If there is a neighbor at the same refinement level, copy from it.
  \item If there is a physical boundary, call the user's routine for
  boundary conditions.
  \item If there is a refinement boundary, call the user's routine for
  refinement boundaries.
\end{itemize}

\begin{figure}
  \centering
  \includegraphics[width=5.0cm]{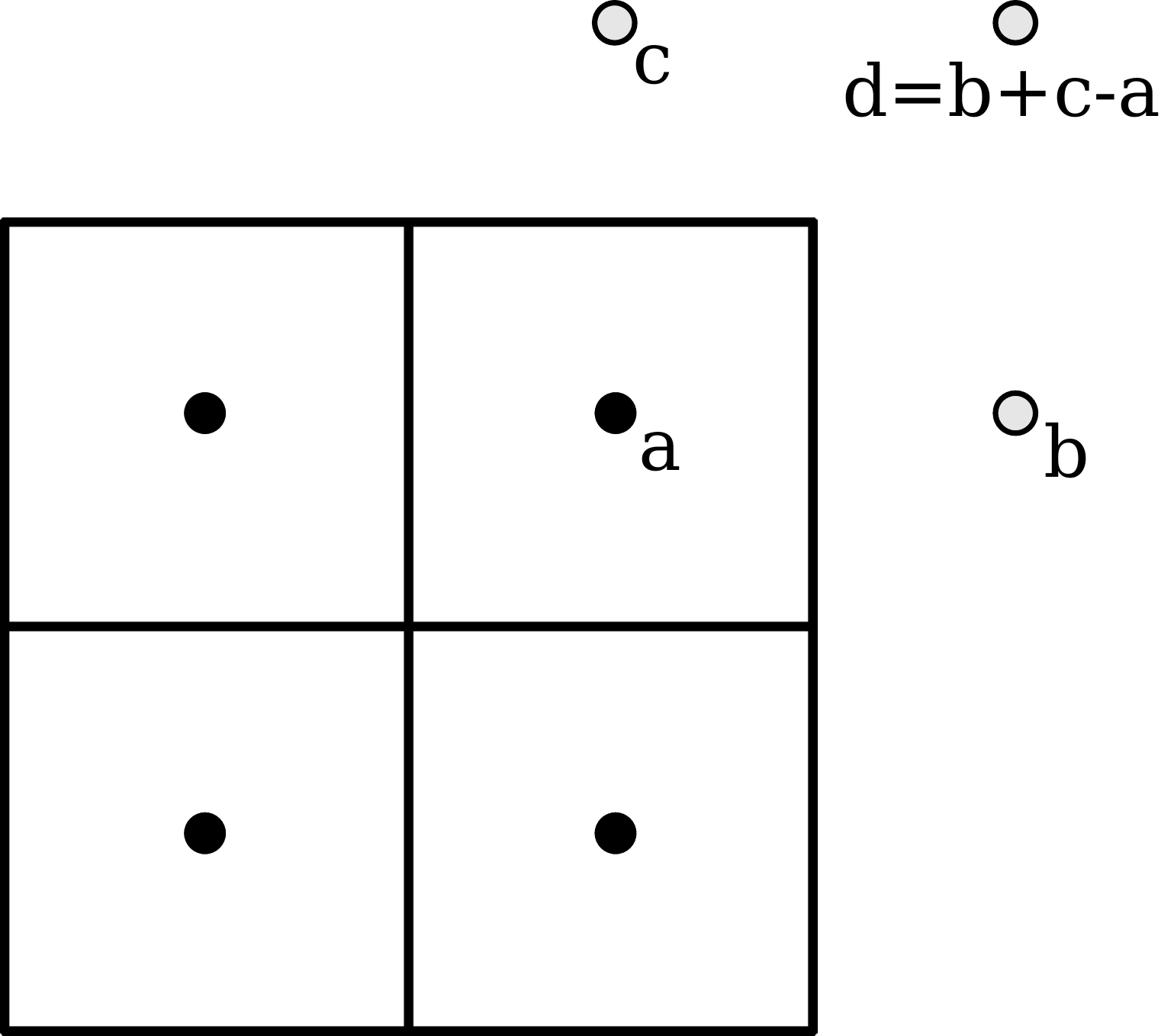} %
  \caption{Illustration of the linear extrapolation procedure for corner ghost
    cells in 2D that is used when the diagonal neighbor is missing. The corner
    ghost cell gets a value $b+c-a$.}
  \label{fig:gc-extrap-2d}
\end{figure}

For the edge and corner ghost cells values are copied if a neighbor at the same
refinement level is present. If there is no such neighbor, for example due to a
physical or refinement boundary, these ghost cells are filled using linear
extrapolation. The extrapolation procedure is illustrated in figure
\ref{fig:gc-extrap-2d}, for a corner ghost cell in 2D. A convenient property of
this approach is that if one later uses bilinear interpolation using the points
$(a,b,c,d)$ the result is equivalent to a linear interpolation based on points
$(a,b,c)$. Furthermore, edge and corner ghost cells can be filled one box at a
time, since they do not depend on ghost cells on neighboring boxes.

Afivo includes procedures to fill ghost cells near refinement boundaries using
linear interpolation. Our approach allows users to construct custom schemes,
which is important because there is no universal `correct' way to do this: one
has to balance higher order (to compensate for the increased discretization
error near the refinement boundary) with conservation principles.

When a second layer of ghost cells is required, we temporarily copy a box to an
enlarged version with two layers of ghost cells, as is also possible in
Paramesh~\cite{Macneice_2000}. In principle, such an enlarged box could also be
used for the first layer of ghost cells, so that no ghost cells need to be
permanently stored. However, then one has to take care not to unnecessarily
recompute ghost values, and extra storage is required to parallelize algorithms.
Our approach of always storing a single layer of ghost cells therefore strikes a
balance between memory usage, simplicity and performance.


\subsection{Prolongation and restriction}
\label{sec:interp-restrict}

In an AMR context, the interpolation of coarse grid values to obtain fine grid
values is often called prolongation. The inverse procedure, namely the averaging
of fine grid values to obtain coarse grid values, is called restriction.

\begin{figure}
  \centering
  \includegraphics[width=4.0cm]{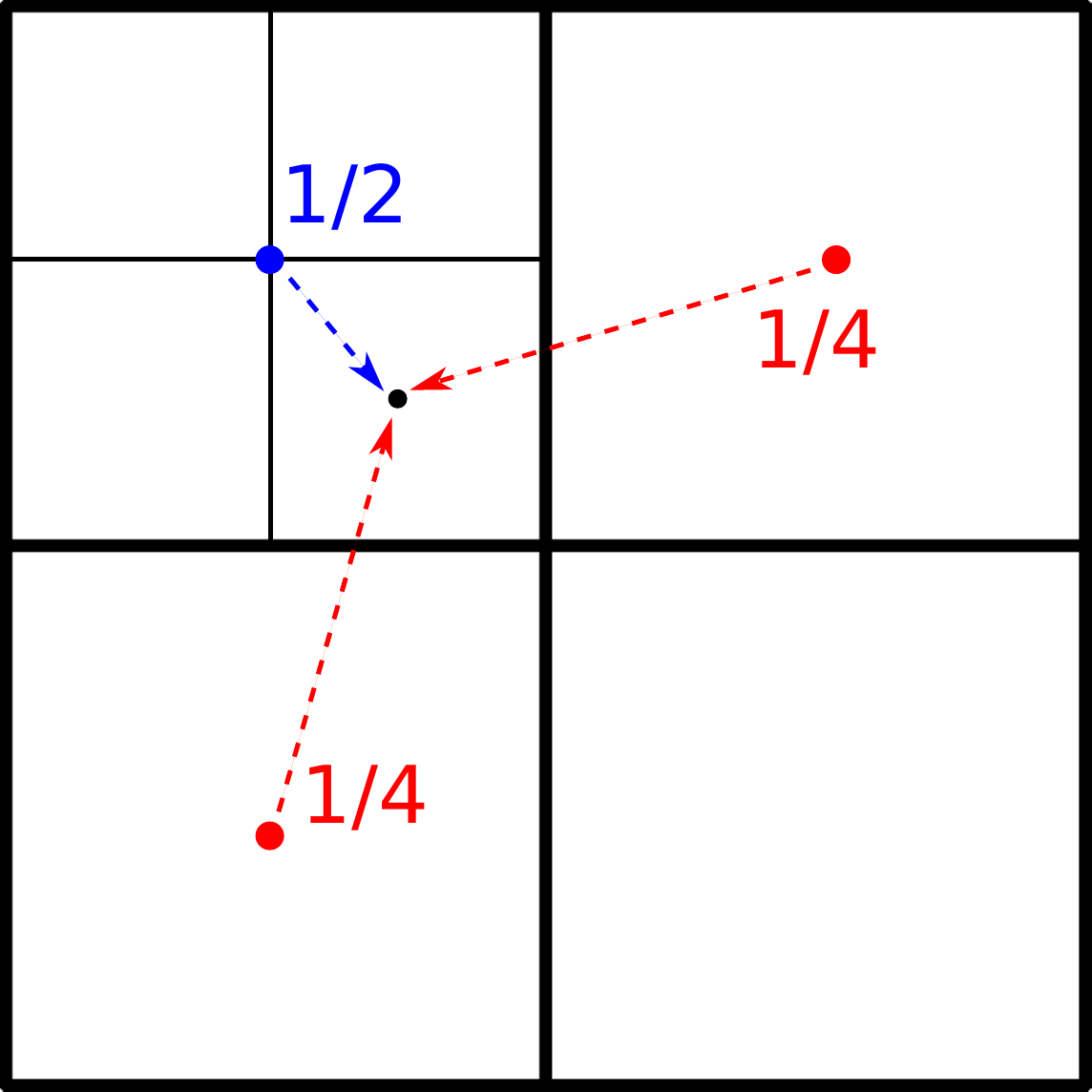} %
  \caption{Schematic drawing of $2-1-1$ interpolation. The three nearest coarse
    grid values are used to interpolate to the center of a fine grid cell. Note
    that the same interpolation scheme can be used for all fine grid cells,
    because of the symmetry in a cell-centered discretization.}
  \label{fig:interp-2d}
\end{figure}

For prolongation, the standard bilinear and trilinear interpolation schemes are
included. Furthermore, schemes with weights
$(\tfrac{1}{2},\tfrac{1}{4},\tfrac{1}{4})$ (2D) and
$(\tfrac{1}{4},\tfrac{1}{4},\tfrac{1}{4},\tfrac{1}{4})$ (3D) are included, as
also implemented in e.g., Boxlib~\cite{www_boxlib}. These linear schemes use
information from the closest and second-closest coarse grid values, thereby
avoiding the use of corner or edge ghost cell values. The 2D case is illustrated
in figure~\ref{fig:interp-2d}. Zeroth-order interpolation, in which the coarse
values are simply copied, is also implemented. The inclusion of higher order and
conservative prolongation methods is left for future work.

As a restriction method Afivo includes averaging, in which the parent gets the
average value of its children. A user can also implement custom interpolation
and restriction methods.


\subsection{OpenMP parallelism}
\label{sec:openmp}

Two conventional methods for parallel computing are OpenMP (shared memory) and
MPI (communicating tasks). Afivo was designed for small scale parallelism, for
example using 16 cores, and therefore only supports OpenMP. Compared to an MPI
implementation, the use of OpenMP has several advantages: data can always be
accessed, sequential (user) code can easily be included, and there is no need
for load balancing or communication between processes. Furthermore, current
systems often have 8 or more CPU cores and tens of gigabytes of RAM, which is
sufficient for many scientific simulations. We remark that for problems
requiring large scale parallelism, there are already a number of MPI-based
frameworks available, see table \ref{tab:amr-frameworks}.

Most operations in Afivo loop over a number of boxes, for example over the
leaves at a certain refinement level. All such loops have been parallelized with
OpenMP. In general, the parallel speedup depends on the cost of the algorithm
that one is using. Because the communication cost (e.g., updating ghost cells)
is always about the same, an expensive algorithm will show a better speedup. On
shared memory systems, it is not unlikely for an algorithm to be memory-bound
instead of CPU-bound.

\subsection{Writing output}
\label{sec:output}

Afivo supports two output formats: VTK unstructured files and Silo files. The
VTK unstructured format can handle much more general grids than quadtree and
octree meshes. This format is therefore computationally more costly to
visualize. Although there is some experimental support for octrees in
VTK~\cite{Carrard_2012}, this support does not yet seem to extend to data
visualization programs such as Paraview \cite{www_paraview} and Visit
\cite{HPV:VisIt}.

Afivo also supports writing Silo files, which include ghost cell information to
prevent gaps in contour or surface plots. These files contain a number of
Cartesian blocks (`quadmeshes' in Silo's terminology). Because writing and
reading a large number of separate blocks is quite costly, we use a simple
algorithm to collect the leaf-boxes (those without children) at a refinement
level into rectangular regions. The algorithm starts with a region $R$ that
consists of a single box. If all the neighbors to the left of $R$ exist, have no
children, and are not yet included in the output, then these neighbors are added
to $R$; otherwise none of them is included. The procedure is repeated in all
directions, until $R$ can no longer grow. Then $R$ is added to the output, and
the procedure starts again until there are no leaf-boxes left.

\section{Multigrid}
\label{sec:multigrid}



Elliptic partial differential equations, such as Poisson's equation, have to be
solved in many applications. Multigrid
methods~\cite{Brandt_2011,Trottenberg_2000_multigrid,Briggs_2000,Hackbusch_1985}
can be used to solve such equations with great efficiency. The error in the
solution is iteratively damped on a hierarchy of grids, with the coarse grids
reducing the low frequency (i.e., long wavelength) error components, and the
fine grids the high frequency components. When using adaptive mesh refinement on
octree grids, geometric multigrid methods have several advantages:
\begin{itemize}
  \item They can run in linear time, i.e., $O(N)$, where $N$ is the number of
  unknowns.
  \item Memory requirements are also linear in the number of unknowns.
  \item The octree already contains the hierarchy of grids required by the
  multigrid method.
  \item Geometric multigrid is \emph{matrix-free}, so that changes in the mesh
  do not incur extra costs (direct methods would have to update their
  factorization).
\end{itemize}
For these reasons, we have implemented a geometric multigrid solver in Afivo,
which can be used to solve problems of the form
\begin{equation}
  A_h(u_h) = \rho_h,
  \label{eq:mg-discr-equation}
\end{equation}
where $A_h$ is a discretized elliptic operator, $\rho_h$ the source term, $u_h$
the solution to be computed and $h$ the mesh spacing. Boundary conditions can be
of Dirichlet, Neumann or periodic type (or a mix of them). A drawback of
geometric multigrid is that the operator $A_h$ also has to be well-defined on
coarse grid levels. This complicates the implementation of e.g., irregular
boundary conditions that do not align with the mesh.

On an octree mesh, the fine grid generally does not cover the whole domain.
Therefore we use Full Approximation Scheme (FAS) version of multigrid, in which
the solution is specified on all levels. The basic multigrid procedures are
summarized below, with a focus on conservative discretization near refinement
boundaries. A general introduction to multigrid methods can be found in
e.g.~\cite{Brandt_2011,Trottenberg_2000_multigrid,Briggs_2000}.

\subsection{Gauss-Seidel red-black smoother}
\label{sec:gsrb}

A \emph{smoother}, which locally smooths out the error in the solution, is a key
component of a multigrid method. Afivo's multigrid module comes with a
collection of so-called Gauss-Seidel red-black smoothers, for Poisson's equation
in 2D, 3D and cylindrical coordinates. These methods operate on one box at a
time, and can do so at any refinement level. How they work is explained below.

Consider an elliptic equation like \eqref{eq:mg-discr-equation} in 2D, using a
5-point numerical stencil. Such an equation relates a value $u_h^{(i,j)}$ at
$(i,j$) to the source term $\rho^{(i,j)}$ and to neighboring values
$u_h^{(i \pm 1,j)}$ and $u_h^{(i,j \pm 1)}$. If the values of the neighbors are
kept fixed, the value $u_h^{(i,j)}$ that locally solves the equation can be
determined. With Gauss-Seidel red-black, such a procedure is applied on a
checkerboard pattern. In two dimensions, points $(i,j)$ can be labeled
\emph{red} when $i+j$ is even and \emph{black} when $i+j$ is odd; the procedure
is analogous for $(i,j,k)$ in 3D. The equation is then first solved for all the
red points while keeping the old black values, and then for the black points.

For example, for Laplace's equation with a standard second order discretization,
a Gauss-Seidel red-black smoother replaces all red points by the average of
their black neighbors, and then vice versa.

\subsection{The V-cycle and FMG-cycle}
\label{sec:mg-v-fmg-cycle}

There exist different multigrid cycles, which control in what order smoothers
(of e.g. Gauss-Seidel red-black type) are used on different grid levels, and how
information is communicated between these levels. The multigrid module of Afivo
implement both the V-cycle and the FMG cycle; both can be called by users.

\subsubsection{V-cycle}

One of the most basic and standard ones is the V-cycle, which is included in
Afivo. This cycle starts at the finest grid, descends to the coarsest grid, and
then goes back up to the finest grid. Consider a grid with levels
$l = 1, 2, \ldots, l_\mathrm{max}$. On each level $l$, $v_h$ denotes the current
approximation to the solution on a grid spacing $h$, and $v_H$ refers to the
(coarse) approximation on level $l-1$ with grid spacing $H=2h$. Furthermore, let
$I_H^h$ be a prolongation operator to go from coarse to fine and $I_h^H$ a
restriction operator to go from fine to coarse, as discussed in section
\ref{sec:interp-restrict}.
The FAS V-cycle can then be described as
\begin{enumerate}
  \item For $l$ from $l_\mathrm{max}$ down to $2$, perform $N_\mathrm{down}$
  (default: two) smoothing steps on level $l$, then compute the residual
  \begin{equation}
  r_h = \rho_h - A_h(v_h).
  \label{eq:residual}
  \end{equation}
  Afterwards update the level $l-1$ coarse grid:
  \begin{enumerate}
    \item Set $v_H \leftarrow I_h^H v_h$, then store a copy $v'_H$ of $v_H$.
    \item Update the coarse grid source term
    \begin{equation}
      \rho_H \leftarrow I_h^H r_h + A_H(v_H).
      \label{eq:coarse-rhs}
    \end{equation}
  \end{enumerate}

  \item \label{step:coarse} Perform $N_\mathrm{base}$ (default: four) relaxation
  steps on level $1$, or apply a direct solver.
  \item For $l$ from $2$ to $l_\mathrm{max}$, perform a
  correction using the data from level $l-1$
  \begin{equation}
    u_h \leftarrow u_h + I_H^h(v_H - v'_H),
    \label{eq:coarse-corr}
  \end{equation}
  then perform $N_\mathrm{up}$ (default: two) relaxation steps on level $l$.
\end{enumerate}




In step~\ref{step:coarse}, relaxation takes place on the coarsest grid. In order
to quickly converge to the solution with a relaxation method, this grid should
contain very few points (e.g., $2\times 2$ or $4\times 4$ in 2D). Alternatively,
a direct solver can be used on the coarsest grid, but such a solver is not yet
included in Afivo. Currently, additional coarse grids are constructed below the
coarsest quadtree/octree level. For example, if a quadtree has boxes of
$16\times 16$ cells, then three coarser levels are added with boxes of
$8\times 8$, $4\times 4$ and $2\times 2$ cells to speed up the multigrid
convergence. Note that a non-square domain will contain more than one
$2\times 2$ box on the coarse grid, and therefore require more coarse grid
relaxation steps.

Successive V-cycles will reduce the residual $r_h$ on the different grid levels,
see the example in section~\ref{sec:mg-convergence}. No automatic error control
has been implemented, so it is up to the user to decide when the residual is
sufficiently small. The residual does typically not need to be reduced to zero,
because the \emph{discretization error} (due to the e.g. second order
discretization) dominates when the residual is small enough. The number of
V-cycles required to reach the discretization error is typically problem(-size)
dependent.



\subsubsection{FMG cycle}

Besides the V-cycle, the full multigrid (FMG) cycle is also implemented in
Afivo. An advantage of the FMG-cycle is that it typically gives convergence up
to the discretization error in one or two iterations. The FMG-cycle operates as
follows:
\begin{enumerate}
  \item If there is no approximate solution yet, set the initial guess to zero
  on all levels, and restrict $\rho$ down to the coarsest grid using $I_h^H$. If
  there is an approximate solution $v$, restrict $v$ down to the coarsest level.
  Use equation~\eqref{eq:coarse-rhs} to set $\rho$ on coarse grids.
  \item For $l = 1, 2, \ldots, l_\mathrm{max}$
  \begin{itemize}
    \item Store the current approximation $v_h$ as $v'_h$.
    \item If $l > 1$, perform a coarse grid correction using equation
    \eqref{eq:coarse-corr}.
    \item Perform a V-cycle starting at level $l$.
  \end{itemize}
\end{enumerate}

\subsection{Conservative filling of ghost cells}
\label{sec:mg-ghost-cells}



\begin{figure}
  \centering
  \includegraphics[width=8cm]{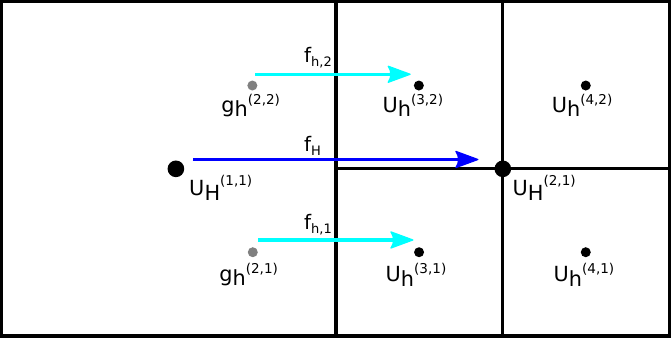}
  \caption{Illustration of a refinement boundary. The cell centers are indicated
    by dots. There are two ghost values (gray dots) on the left of the refinement
    boundary. Fluxes across the refinement boundary are indicated by arrows.}
  \label{fig:mg-ref-bound}
\end{figure}

As discussed in section \ref{sec:ghost-cell}, ghost cells are used to facilitate
computations near refinement boundaries. How these ghost cells are filled
affects the multigrid solution and convergence behavior. In Afivo, we have
implemented \emph{conservative} schemes for filling ghost cells
\cite{Bai_1987,Trottenberg_2000_multigrid}. A conservative scheme ensures that
the coarse flux across a refinement boundary equals the average of the fine
fluxes, see figure \ref{fig:mg-ref-bound}. To illustrate why a conservative
discretization is important, consider an equation of the form
$\nabla \cdot \vec{F} = \rho$. The divergence theorem gives
\begin{equation}
  \int_V \rho \, dV = \int_V \nabla \cdot \vec{F} \, dV = \int \vec{F} \cdot
  \vec{n} \, dS,
\end{equation}
where the last integral runs over the surface of the volume $V$, and $\vec{n}$
is the normal vector to this surface. When the fine and coarse fluxes are
consistent, the integral over $\rho$ will be same on the fine and the coarse
grid.

The construction of a conservative scheme for filling ghost cells is perhaps
best explained with an example.
Consider a 2D Poisson problem
\begin{equation*}
  \nabla^2 u = \nabla \cdot (\nabla u) = \rho.
\end{equation*}
With a standard 5-point stencil for the Laplace operator
the coarse flux $f_H$ across the refinement boundary in figure
\ref{fig:mg-ref-bound} is given by
\begin{equation*}
  f_H = [u_H^{(2,1)} - u_H^{(1,1)}]/H,
\end{equation*}
and on the fine grid, the two fluxes are given by
\begin{align*}
  f_{h,1} &= [u_h^{(3,1)} - g_h^{(2,1)}]/h,\\
  f_{h,2} &= [u_h^{(3,2)} - g_h^{(2,2)}]/h.
\end{align*}
The task is now to fill the ghost cells $g_h^{(2,1)}$ and $g_h^{(2,2)}$ in such
a way that the coarse flux equals the average of the fine fluxes:
\begin{equation}
  \label{eq:mg-flux-cons}
  f_H = (f_{h,1} + f_{h,2})/2.
\end{equation}
To relate $u_H^{(2,1)}$ to the fine-grid values $u_h$, the restriction operator
$I_h^H$ needs to be specified.
In our implementation, this operator does averaging over the children.
The constraint from equation \eqref{eq:mg-flux-cons} can then be written as
\begin{equation}
  g_h^{(2,1)} + g_h^{(2,2)} = u_H^{(1,1)} + \frac{3}{4} \left(u_h^{(3,1)} + u_h^{(3,2)}\right)
  - \frac{1}{4} \left(u_h^{(4,1)} + u_h^{(4,2)}\right).
  \label{eq:mg-gc-condition}
\end{equation}
Any scheme for the ghost cells that satisfies this constraint leads to a
conservative discretization.

Bilinear \emph{extrapolation} (similar to standard bilinear interpolation) satisfies equation~\eqref{eq:mg-gc-condition} and gives
the following scheme for $g_h^{(2,1)}$
\begin{equation*}
  g_h^{(2,1)} = \frac{1}{2} u_H^{(1,1)} + \frac{9}{8} u_h^{(3,1)} -
  \frac{3}{8} \left (u_h^{(3,2)} + u_h^{(4,1)} \right)
  + \frac{1}{8} u_h^{(4,2)}.
\end{equation*}
(The scheme for $g_h^{(2,2)}$ follows from symmetry.) Another option is to use
only the closest two neighbors for the extrapolation, which gives the following
expression for $g_h^{(2,1)}$
\begin{equation*}
  g_h^{(2,1)} = \frac{1}{2} u_H^{(1,1)} + u_h^{(3,1)} -
  \frac{1}{4} \left (u_h^{(3,2)} + u_h^{(4,1)} \right).
\end{equation*}
This last scheme is how ghost cells at refinement boundaries are filled by
default in Afivo. In three dimensions, the scheme becomes
\begin{equation*}
  g_h^{(2,1,1)} = \frac{1}{2} u_H^{(1,1,1)} + \frac{5}{4} u_h^{(3,1,1)} -
  \frac{1}{4} \left (u_h^{(4,1,1)} + u_h^{(3,2,1)} + u_h^{(3,1,2)}\right).
\end{equation*}

We have observed that filling ghost cells as described above can reduce the
multigrid convergence rate, in particular in 3D. There are two reasons: first, a
type of local extrapolation is performed, and the larger the coefficients in
this extrapolation are, the more smoothing is required to reduce errors. Second,
cells near a refinement boundary do not locally solve the linear equation after
an Gauss-Seidel red-black update, if one takes into account that the ghost cells
also have to be updated. It is possible to fix this, in a similar way as one can
change the stencil near physical boundaries instead of using ghost cells, but
near a `refinement corner' the situation is more complicated.

\subsection{Including discontinuities in $\varepsilon$}
\label{sec:mg-varepsilon}


For the more general equation $\nabla \cdot (\varepsilon \nabla \phi) = \rho$ we
have implemented a special case: $\varepsilon$ jumps from $\varepsilon_1$ to
$\varepsilon_2$ at a cell face. Local reconstruction of the solution shows that
the flux through the cell face is then given by
\begin{equation}
  \frac{2 \, \varepsilon_1 \varepsilon_2}{\varepsilon_1 + \varepsilon_2} \, \frac{\phi_{i+1} -
    \phi_i} {h}.
\end{equation}
In other words, the flux is multiplied by the harmonic mean of the
$\varepsilon$'s (see e.g., chapter 7.7 of \cite{Trottenberg_2000_multigrid}).
The ghost cell schemes described above for constant $\varepsilon$ still ensure
flux conservation, because the coarse and fine flux are multiplied by the same
factor. The jump should occur at a cell face at \emph{all refinement levels},
which is equivalent to requiring that it occurs at a coarse grid cell face.

\subsection{Supported operators}
\label{sec:mg-operators}

The following elliptic operators have been implemented in Afivo:
\begin{itemize}
  \item 2D/3D Laplacian in Cartesian coordinates, using a 5 and 7-point stencil
  respectively.
  \item 2D/3D Laplacian with a jump in coefficient on a cell face, as discussed
  in the previous section. A custom prolongation (interpolation) method that
  uses the locally reconstructed solution is also included.
  \item Cylindrical Laplacian in $(r,z)$-coordinates, also supporting a jump in
  coefficient on a cell face.
\end{itemize}
Furthermore, a Laplacian with support for internal boundaries has been
implemented, which makes use of a level set function to determine the location
of the boundaries. At the moment, this only works if the boundary can also be
resolved on the coarse grid. The future implementation of a direct sparse method
for the coarse grid equations will enable this functionality more generally,
because the coarse grid can then have a higher resolution.

Users can also define custom elliptic operators, as well as custom smoothers and
prolongation and restriction routines. One of the examples included with Afivo
shows how the diffusion equation $\partial_t n = D \nabla^2 n$ can be solved
with a backward Euler scheme by defining such a custom operator.

\section{Examples}
\label{sec:examples}

Several examples that demonstrate how to use Afivo are included in the
\texttt{examples} folder of Afivo's source code, see section
\ref{sec:afivo-basics}. Here we discuss a few of the examples in detail.

\subsection{Multigrid convergence}
\label{sec:mg-convergence}

\begin{figure*}
  \centering
  \includegraphics[width=6cm]{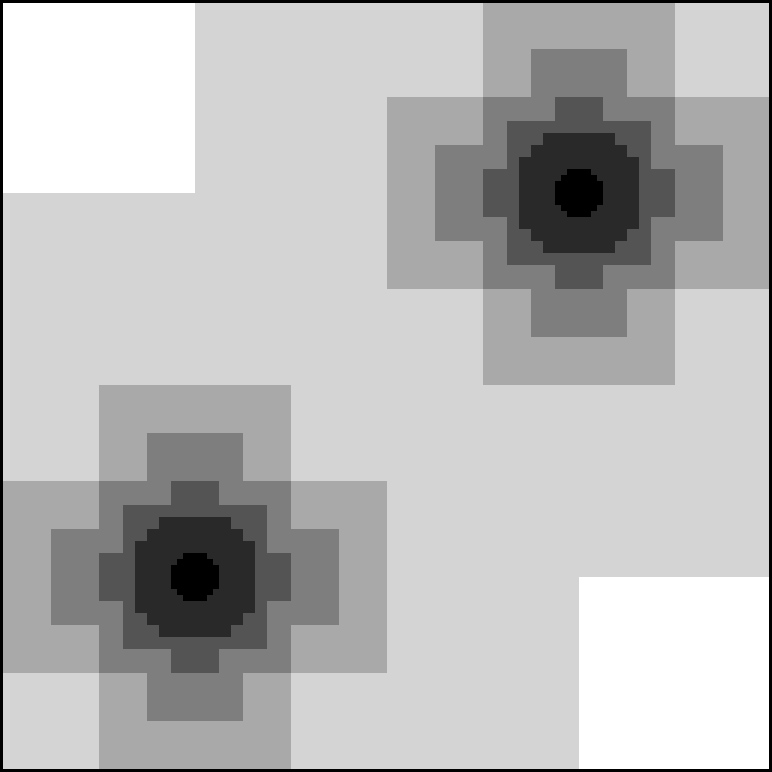} %
  \hspace{1cm}
  \includegraphics{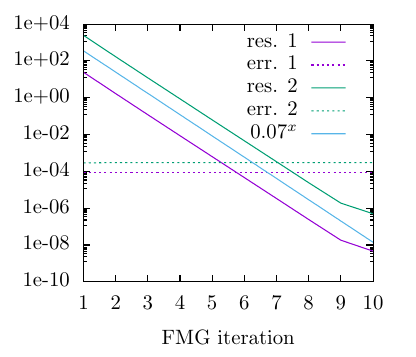}
  \caption{Left: mesh spacing used for the multigrid examples, in a
    $[0,1] \times [0,1]$ domain. Black indicates $\Delta x = 2^{-11}$ and white
    $\Delta x = 2^{-5}$. Right: the maximum residual and maximum error versus
    FMG iteration. Case 1 corresponds to a standard Laplacian (equation
    \eqref{eq:mg-example-lpl-1}) and case 2 to the cylindrical case with a jump
    in $\varepsilon$ (equation \eqref{eq:mg-example-lpl-2}).}
  \label{fig:mg-ex1}
\end{figure*}

In this section we present two test problems to demonstrate the multigrid
behavior on a partially refined mesh. We use the method of manufactured
solutions: from an analytic solution the source term and boundary
conditions are computed. Two test problems are considered, a
constant-coefficient Poisson equation in 2D
\begin{equation}
  \nabla^2 u = \nabla \cdot (\nabla u) = \rho
  \label{eq:mg-example-lpl-1}
\end{equation}
and a problem with cylindrical symmetry in $(r,z)$ coordinates
\begin{equation}
  \frac{1}{r} \partial_r (r \varepsilon \partial_r u) +
  \partial_z (\varepsilon \partial_z u) = \rho,
  \label{eq:mg-example-lpl-2}
\end{equation}
both on a two-dimensional rectangular domain $[0,1] \times [0,1]$.
For the second case, $\varepsilon$ has a value of $100$ in the lower left
quadrant $[0,0.25] \times [0,0.25]$, and a value of $1$ in the rest of the domain.
In both cases, we pick the following solution for $u$
\begin{equation}
  u(r) = \exp(\vectornorm{\vec{r}-\vec{r}_1}/\sigma) + \exp(\vectornorm{\vec{r}-\vec{r}_2}/\sigma),
\end{equation}
where $\vec{r_1} = (0.25, 0.25)$, $\vec{r_2} = (0.75, 0.75)$ and
$\sigma = 0.04$. An analytic expression for the source term $\rho$ is obtained
by plugging the solution in equations \eqref{eq:mg-example-lpl-1} and
\eqref{eq:mg-example-lpl-2} (note that jumps in $\varepsilon$ also contribute to
the source term $\rho$). The solution is used to set Dirichlet boundary
conditions. For these examples, we have used $N_\mathrm{down} = 2$,
$N_\mathrm{up} = 2$ and $N_\mathrm{base} = 4$ smoothing steps, and boxes with
$8^2$ cells.

The refinement criterion is based on the source term $\rho$: refine if
$\Delta x^2 |\rho| / \varepsilon > 10^{-3}$, where $\varepsilon$ is one for the
first problem. The resulting mesh spacing, which is the same for both problems,
is shown in figure \ref{fig:mg-ex1}a. Figure~\ref{fig:mg-ex1}b shows that in
both cases, one FMG (full multigrid) cycle is enough to achieve convergence up
to the discretization error.
Consecutive FMG cycles further reduce the residual $r = \rho - \nabla^2 u$. The
convergence behavior is similar for both cases, with each iteration reducing
the residual by a factor of about $0.07$. This factor decreases when more
smoothing steps are taken and when a higher order prolongation or restriction
method is used. For this example we have used first order prolongation and
simple averaging for restriction, as discussed in section
\ref{sec:interp-restrict}. The offset between the lines is caused by the
$\varepsilon = 100$ region, which locally amplifies the source term by a factor
of 100.

\subsection{Multigrid performance and scaling}
\label{sec:mg-scaling}

\begin{figure*}
  \centering
  \includegraphics[width=7cm]{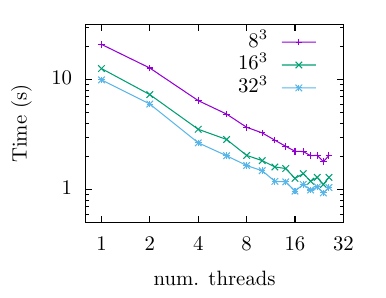}
  \hspace{1cm}
  \includegraphics[width=7cm]{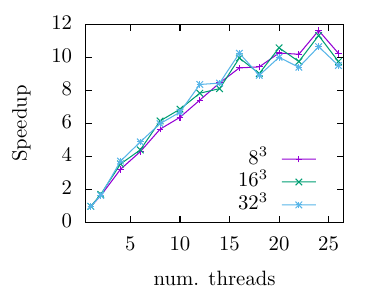}
  \caption{Duration (left) and speedup (right) of a single FMG cycle on a
    uniformly refined grid of $512^3 \approx 134 \times 10^6$ cells versus the
    number of OpenMP threads. Results are shown for octrees with boxes of $8^3$,
    $16^3$ and $32^3$ cells.}
  \label{fig:mg-scaling}
\end{figure*}

Here we briefly investigate the performance and scaling of the multigrid
routines. Although the numbers presented here depend on the particular system
and compiler used, they can be used to estimate feasible problem sizes with
Afivo. As a performance test we use a 3D Poisson problem
$$\nabla^2 \phi = 1,$$
on a domain $[0,1]^3$ with $\phi = 0$ at the boundaries. For simplicity (and for
comparison with other methods), the domain is uniformly refined up to a
resolution of $512^3$ cells.

Figure \ref{fig:mg-scaling} shows the duration of a single FMG cycle versus the
number of processor cores used, again using $N_\mathrm{down} = 2$,
$N_\mathrm{up} = 2$ and $N_\mathrm{base} = 4$ smoothing steps. Curves are shown
for box sizes of $8^3$, $16^3$ and $32^3$, which affect the overhead of the
adaptive octree mesh. The runs were performed on a node with Xeon E5-2680
processors ($2.8 \, \textrm{GHz}$, 20 cores per node).

The maximal speedups are about $4.9$ ($8^3$ case), $5.4$ ($16^3$ case) and $5.7$
($32^3$ case). The performance of the geometric multigrid algorithm, which
performs only a few additions and multiplications per cell during each smoothing
step, is probably bound by the memory bandwidth of the system. We also see that
performance is increased when using larger boxes, because this reduces the
overhead due to the filling of ghost cells. For the $32^3$ case with 16 cores,
the time spent per unknown is about $10 \, \textrm{ns}$, whereas it is about
$22 \, \textrm{ns}$ for the $8^3$ case with 16 cores.

\subsection{Discharge model}
\label{sec:example-discharge}

In previous studies~\cite{Nijdam_Teunissen_2016,teunissen_2017_afivo_streamer},
Afivo has already been used to study the guiding of so-called streamer
discharges in 3D. For simplicity, we here consider a simpler 2D plasma fluid
model for electric gas discharges~\cite{Luque_2012}. This model is used to
simulate the destabilization of a planar ionization wave in pure nitrogen, in a
background field above the breakdown threshold. The destabilization of such
planar ionization waves has been investigated mainly analytically in the
past~\cite{Arrays_2004,Derks_2008,Ebert_nonlin_2010}.

The model is kept as simple as possible: it contains only electrons and positive
ions, no photo-ionization and no plasma chemistry. The evolution of the electron
and ion density ($n_e$ and $n_i$) is then described by the following equations:
\begin{align}
  \label{eq:fluid-model}
  \partial_t n_e &= \nabla \cdot (\mu_e \vec{E} n_e + D_e \nabla n_e) + \alpha(E) \mu_e E n_e,\\
  \partial_t n_i &= \alpha(E) \mu_e E n_e,\\
  \nabla^2 \phi &=  - e (n_i - n_e)/\varepsilon_0,\qquad \vec{E} = -\nabla\phi,
\end{align}
where $\mu_e$ is the electron mobility, $D_e$ the electron diffusion
coefficient, $\alpha(E)$ the ionization coefficient, $\vec{E}$ the electric
field, $\phi$ the electrostatic potential,
$\varepsilon_0$ the permittivity of vacuum and $e$ the elementary charge.
The motion of ions is not taken into account here. The electrostatic potential
is computed with the FMG multigrid routine described in section
\ref{sec:mg-v-fmg-cycle}. The electric field at cell faces is then calculated by
taking central differences.

For simplicity, we use a constant mobility
$\mu_e = 0.03 \, \textrm{m}^2/(\mathrm{Vs})$, a constant diffusion coefficient
$D_e = 0.2 \, \textrm{m}^2/\mathrm{s}$ and we take an analytic expression for
the ionization coefficient
$\alpha(E) = \exp\left[10.4 + 0.601 \log(E/E^*) - 186 (E^*/E)\right]$, with
$E^* = 1 \, \textrm{kV/cm}$~\cite{Li_hybrid_ii_2012}. These coefficients roughly
correspond to nitrogen at room temperature and normal pressure. In a more
realistic model, one would typically include tabulated transport coefficients to
make the results more realistic. Such coefficients can be computed with a
Boltzmann solver (e.g., \cite{Hagelaar_2005,Dujko_2011}) or particle swarms
(e.g., \cite{Li_hybrid_i_2010,Rabie_2016}).

The electron flux is computed as in \cite{Montijn_2006}. The diffusive part is
computed using central differences and the drift part is computed using the
Koren limiter \cite{koren_limiter}. The Koren limiter was not designed to
include refinement boundaries, and we use linear interpolation to obtain
fine-grid ghost values. These ghost cells lie inside a coarse-grid neighbor
cell, and we limit them to twice the coarse values to preserve positivity.

Time stepping is also performed as in \cite{Montijn_2006}, using the explicit
trapezoidal rule. The global time step is taken as the minimum over the cells of
\begin{itemize}
  \item CFL condition:
  $\frac{1}{2} / \left(|v_x| / \Delta x + |v_y| / \Delta x \right)$, where $v_x$
  and $v_y$ are the $x$ and $y$-component of the electron drift
  \item Explicit diffusion limit: $\Delta x^2 / (4 D_e)$
  \item Dielectric relaxation time: $\varepsilon_0 / (e \mu_e n_e)$
\end{itemize}

The refinement criterion is based on the ionization coefficient $\alpha$, which
depends on the local electric field. The reasoning behind this is that
$1/\alpha$ is a typical length scale for the electron and ion density gradients
and the width of space charge layers~\cite{Ebert_nonlin_2010}. Where
$n_e > 1 \, \textrm{m}^{-3}$ (an arbitrary small value) and
$\alpha \Delta x > 0.8$, the mesh is marked for refinement. Elsewhere the mesh
is marked for derefinement when $\Delta x < 25 \, \mu\textrm{m}$ and
$\alpha \Delta x < 0.1$. The quadtree mesh for this example was constructed from
boxes containing $8^2$ cells.

\begin{figure*}
  \centering
  \includegraphics[width=\textwidth]{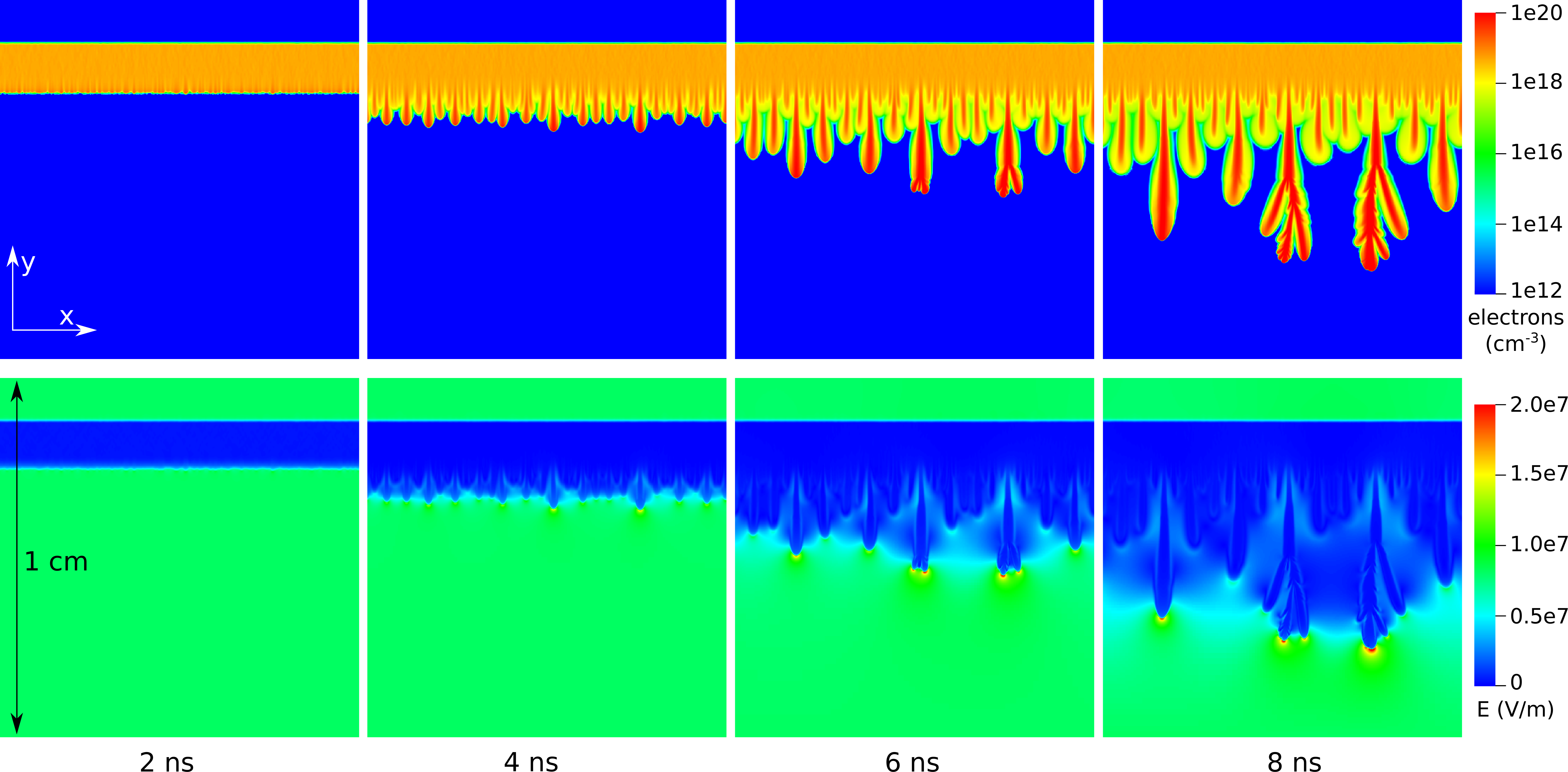}
  \caption{The evolution of the electron density (top) and electric field
    (bottom) in a 2D electric discharge simulation in nitrogen at standard
    temperature and pressure. The discharge started from a pre-ionized layer,
    which destabilizes into streamer channels. A zoom-in of the mesh around a
    streamer head at $t = 8 \, \textrm{ns}$ is shown in
    figure~\ref{fig:ex-streamer-zoom-mesh}.}
  \label{fig:ex-streamer-dns-fld}
\end{figure*}

The model described above is used to simulate discharges in a domain of
$(1 \, \textrm{cm})^2$, see figure \ref{fig:ex-streamer-dns-fld}. Initially, a
density $n_0$ of approximately $10^{15} \, \textrm{cm}^{-3}$ electrons and ions
is present between $y = 9 \, \textrm{mm}$ and $y = 9.5 \, \textrm{mm}$,
elsewhere the density is zero. The precise density in each cell is drawn using
random numbers, by taking samples from a normal distribution with mean and
variance $n_0 \Delta x^3$, with $\Delta x \approx 9.8 \mu\mathrm{m}$ in the
region of the initial condition. For $n_0 \Delta x^3 \gg 1$, as we have here,
this approximates the Poisson distribution of physical particle noise (when the
simulation would be truly 3D). At $y = 1 \, \textrm{cm}$ the domain is grounded,
and at $y = 0$ a background field of $8 \, \textrm{MV/m}$ is applied through a
Neumann condition for the electric potential; therefore the electrons drift
downward in the field. The electron and ion density at the $y$-boundaries are
set to zero, and the domain has periodic boundary conditions in the
$x$-direction.

\begin{figure*}
  \centering
  \includegraphics[width=\textwidth]{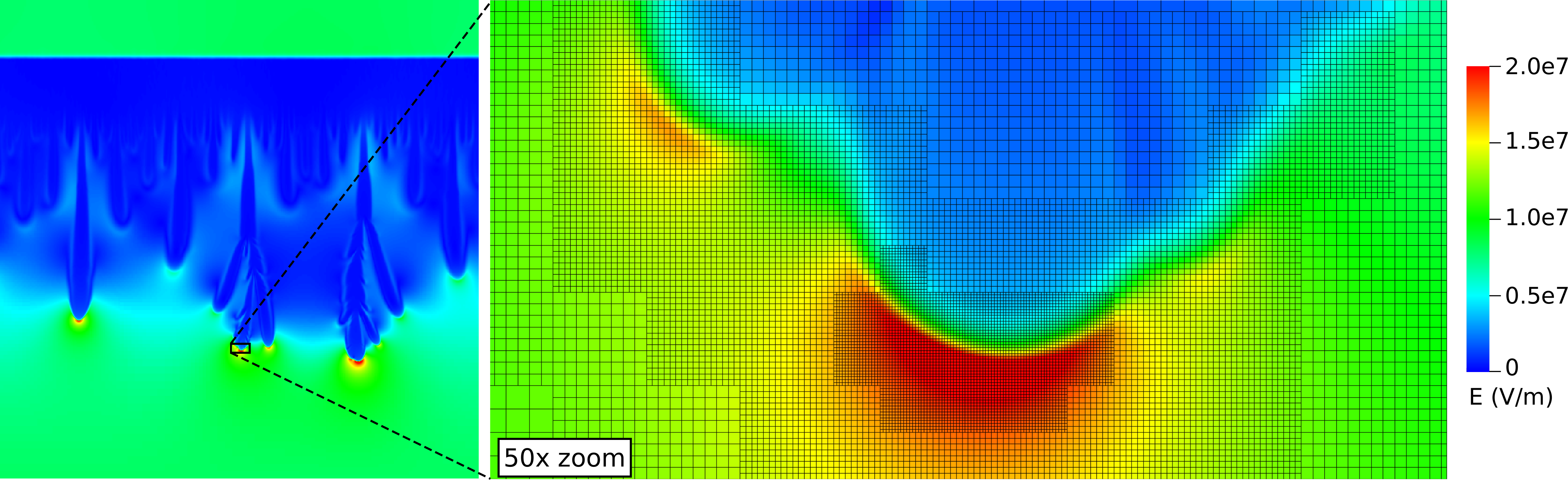}
  \caption{The full domain and a 50 times zoom, which shows the electric field
    and the mesh around a streamer head at $t = 8 \, \textrm{ns}$. The finest
    grid spacing is $\Delta x \approx 1.22 \, \mu\textrm{m}$.
  }
  \label{fig:ex-streamer-zoom-mesh}
\end{figure*}

Figure \ref{fig:ex-streamer-dns-fld} shows how the electron density and the
electric field evolve in time. At first, the pre-ionized layer grows rather
homogeneously downwards due to electron drift and creation of more charges
through impact ionization, but because small inhomogeneities locally enhance the
electric field~\cite{Ebert_nonlin_2010}, the layer quickly destabilizes into
streamer channels. The faster channels electrically screen the slower ones,
reducing the number of active channels over time. Figure
\ref{fig:ex-streamer-zoom-mesh} shows a zoom of the adaptively refined mesh at
$t = 8 \, \textrm{ns}$.

\subsection{Toy model of particles interacting through gravity}
\label{sec:gravity-ex}

Afivo includes basic functionality for particle simulations. A bi/tri-linear
interpolation procedure is provided to interpolate fields at particle positions.
There is also a routine for mapping a list of particle coordinates and
corresponding weights to densities on a grid. Particles can be assigned to the
nearest cell center, or a cloud-in-cell shape function~\cite{Hockney_1988} can
be used\footnote{Near refinement boundaries, we revert to the nearest cell to
  preserve the total particle density.}.

To demonstrate the particle coupling, we present results of a simple toy model
for self-gravitating particles in a fully periodic domain. The model is inspired
by N-body codes for gravitating systems~\cite{Dehnen_2011}. Here we do not take
the short-range interaction between particles into account, and the model does
not strictly conserve energy. For simplicity, we omit all units in the model's
description below and we set $4 \pi G = 1$, where $G$ is the gravitational
constant.

\begin{figure*}
  \centering
  \includegraphics[width=\textwidth]{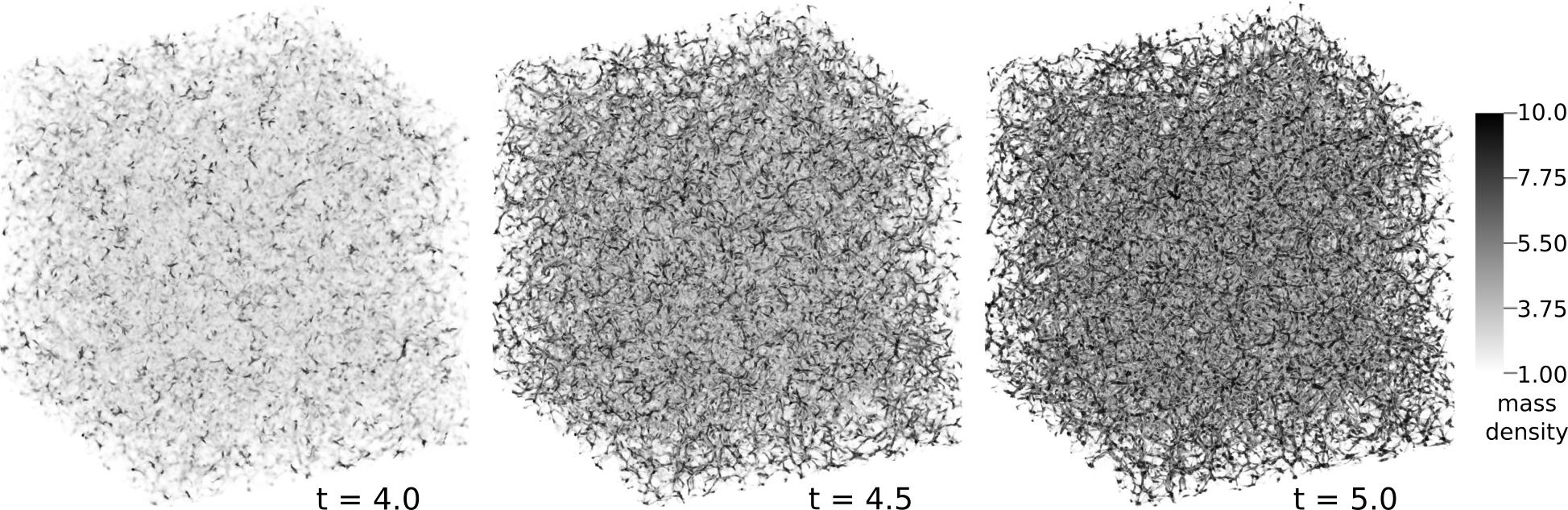}
  \caption{Evolution of the mass density in a 3D periodic system with $10^8$
    particles interacting through gravity. Initially, the particles were
    uniformly distributed. The visualization was made with Visit
    \cite{HPV:VisIt}, using volume rendering.}
  \label{fig:gravity}
\end{figure*}

Initially, $10^8$ particles are uniformly distributed over a unit cube, using
pseudorandom numbers. The initial velocities are set to zero. Each particle has
a mass of $10^{-8}$, so that the mean mass density is one. At each time step,
particle positions and velocities are updated using a synchronized leapfrog
scheme:
\begin{align*}
  \mathbf{x}_{t+1/2} &= \mathbf{x}_{t} + \tfrac{1}{2} \Delta t \, \mathbf{v}_{t},\\
  \mathbf{v}_{t+1} &= \mathbf{v}_{t} + \tfrac{1}{2} \Delta t \, \mathbf{g}_{t+1/2},\\
  \mathbf{x}_{t+1} &= \mathbf{x}_{t+1/2} + \tfrac{1}{2} \Delta t \, \mathbf{v}_{t+1}.
\end{align*}
The gravitational acceleration $\mathbf{g}_{t+1/2}$ is computed by central
differencing of the gravitational potential
$\mathbf{g}_{t+1/2} = -\nabla \phi_{t+1/2}$, and $\phi$ is obtained by solving
Poisson's equation
\begin{equation*}
  \nabla^2 \phi_{t+1/2} = \rho_{t+1/2} - \bar{\rho},\\
\end{equation*}
where $\rho_{t+1/2}$ is the mass density at $t + 1/2$. The mean mass density
$\bar{\rho}$ is subtracted to ensure a fully periodic solution exists, as it
follows from the divergence theorem that the integrated source term has to be
zero.

During the simulation, the mesh is refined where cells contain more than 100
simulation particles, and refinement is removed when boxes contain less than 4
particles. At most seven refinement levels are used, so that the finest grid has
a spacing of about $2 \cdot 10^{-3}$. A constant time step $\Delta t = 10^{-2}$
is used. Figure~\ref{fig:gravity} shows the evolution of the mass density up to
$t = 5$. Small fluctuations in the initial particle density grow over time, and
eventually dense and dilute regions form a complex structure. Up to $t = 3$, the
domain contains about $2$ million cells, but as more and more fine-scale
structure forms, about $10^8$ cells are used at $t = 5$.

\section{Conclusion \& outlook}
\label{sec:conclusion-outlook}

This paper describes Afivo, a framework for parallel simulations on adaptively
refined quadtree/octree grids with a geometric multigrid solver. We have tried
to keep the framework simple to facilitate modification, so it can be used to
experiment with AMR algorithms and methods. An overview of Afivo's main data
structures and procedures was given, and the included geometric multigrid
solvers have been described. We have presented examples of the multigrid
convergence and scaling, of a simplified discharge model in 2D, and of a toy
model for gravitationally interacting particles in 3D.

Future developments will focus on the inclusion of a sparse direct solver that
can handle the coarse grid of the multigrid procedure. This will make it easier
to include irregular boundary conditions in the multigrid solver, to enable for
example the inclusion of curved electrodes in electrostatic calculations.

{\bf Acknowledgments} We thank Margreet Nool for her help with the documentation
and the examples. While developing Afivo, JT was supported by project 10755 of
the Dutch Technology Foundation STW, which is part of the Netherlands
Organisation for Scientific Research (NWO), and which is partly funded by the
Ministry of Economic Affairs. JT is now supported by postdoctoral fellowship
12Q6117N from Research Foundation -- Flanders (FWO).

\bibliographystyle{model1a-num-names}
\bibliography{big_bib_170116}

\begin{thebibliography}{40}
\expandafter\ifx\csname natexlab\endcsname\relax\def\natexlab#1{#1}\fi
\providecommand{\url}[1]{\texttt{#1}}
\providecommand{\href}[2]{#2}
\providecommand{\path}[1]{#1}
\providecommand{\DOIprefix}{doi:}
\providecommand{\ArXivprefix}{arXiv:}
\providecommand{\URLprefix}{URL: }
\providecommand{\Pubmedprefix}{pmid:}
\providecommand{\doi}[1]{\href{http://dx.doi.org/#1}{\path{#1}}}
\providecommand{\Pubmed}[1]{\href{pmid:#1}{\path{#1}}}
\providecommand{\bibinfo}[2]{#2}
\ifx\xfnm\relax \def\xfnm[#1]{\unskip,\space#1}\fi
\bibitem[{Koren(1993)}]{koren_limiter}
\bibinfo{author}{B.~Koren}, in: \bibinfo{editor}{C.~Vreugdenhil},
  \bibinfo{editor}{B.~Koren} (Eds.), \bibinfo{booktitle}{Numerical Methods for
  Advection-Diffusion Problems}, \bibinfo{publisher}{Braunschweig/Wiesbaden:
  Vieweg}, \bibinfo{year}{1993}, pp. \bibinfo{pages}{117--138}.
\bibitem[{Vitello et~al.(1994)Vitello, Penetrante, and Bardsley}]{Vitello_1994}
\bibinfo{author}{P.~A. Vitello}, \bibinfo{author}{B.~M. Penetrante},
  \bibinfo{author}{J.~N. Bardsley}, \bibinfo{journal}{Physical Review E}
  \bibinfo{volume}{49} (\bibinfo{year}{1994}) \bibinfo{pages}{5574–5598}.
  \URLprefix \url{http://dx.doi.org/10.1103/PhysRevE.49.5574}.
  \DOIprefix\doi{10.1103/physreve.49.5574}.
\bibitem[{Yi and Williams(2002)}]{Yi_2002}
\bibinfo{author}{W.~J. Yi}, \bibinfo{author}{P.~F. Williams},
  \bibinfo{journal}{J. Phys. D: Appl. Phys.} \bibinfo{volume}{35}
  (\bibinfo{year}{2002}) \bibinfo{pages}{205--218}. \URLprefix
  \url{http://dx.doi.org/10.1088/0022-3727/35/3/308}.
  \DOIprefix\doi{10.1088/0022-3727/35/3/308}.
\bibitem[{Ebert et~al.(2010)Ebert, Nijdam, Li, Luque, Briels, and van
  Veldhuizen}]{Ebert_2010}
\bibinfo{author}{U.~Ebert}, \bibinfo{author}{S.~Nijdam},
  \bibinfo{author}{C.~Li}, \bibinfo{author}{A.~Luque},
  \bibinfo{author}{T.~Briels}, \bibinfo{author}{E.~van Veldhuizen},
  \bibinfo{journal}{Journal of Geophysical Research} \bibinfo{volume}{115}
  (\bibinfo{year}{2010}). \URLprefix
  \url{http://dx.doi.org/10.1029/2009JA014867}.
  \DOIprefix\doi{10.1029/2009ja014867}, \bibinfo{note}{a00E43}.
\bibitem[{Pancheshnyi et~al.(2008)Pancheshnyi, S{\'e}gur, Capeill{\`e}re, and
  Bourdon}]{Pancheshnyi_2008}
\bibinfo{author}{S.~Pancheshnyi}, \bibinfo{author}{P.~S{\'e}gur},
  \bibinfo{author}{J.~Capeill{\`e}re}, \bibinfo{author}{A.~Bourdon},
  \bibinfo{journal}{Journal of Computational Physics} \bibinfo{volume}{227}
  (\bibinfo{year}{2008}) \bibinfo{pages}{6574--6590}. \URLprefix
  \url{http://dx.doi.org/10.1016/j.jcp.2008.03.020}.
  \DOIprefix\doi{10.1016/j.jcp.2008.03.020}.
\bibitem[{Montijn et~al.(2006)Montijn, Hundsdorfer, and Ebert}]{Montijn_2006}
\bibinfo{author}{C.~Montijn}, \bibinfo{author}{W.~Hundsdorfer},
  \bibinfo{author}{U.~Ebert}, \bibinfo{journal}{Journal of Computational
  Physics} \bibinfo{volume}{219} (\bibinfo{year}{2006})
  \bibinfo{pages}{801--835}. \URLprefix
  \url{http://dx.doi.org/10.1016/j.jcp.2006.04.017}.
  \DOIprefix\doi{10.1016/j.jcp.2006.04.017}.
\bibitem[{Li et~al.(2012)Li, Ebert, and Hundsdorfer}]{Li_hybrid_ii_2012}
\bibinfo{author}{C.~Li}, \bibinfo{author}{U.~Ebert},
  \bibinfo{author}{W.~Hundsdorfer}, \bibinfo{journal}{Journal of Computational
  Physics} \bibinfo{volume}{231} (\bibinfo{year}{2012})
  \bibinfo{pages}{1020--1050}. \URLprefix
  \url{http://dx.doi.org/10.1016/j.jcp.2011.07.023}.
  \DOIprefix\doi{10.1016/j.jcp.2011.07.023}.
\bibitem[{Luque and Ebert(2012)}]{Luque_2012}
\bibinfo{author}{A.~Luque}, \bibinfo{author}{U.~Ebert},
  \bibinfo{journal}{Journal of Computational Physics} \bibinfo{volume}{231}
  (\bibinfo{year}{2012}) \bibinfo{pages}{904--918}. \URLprefix
  \url{http://dx.doi.org/10.1016/j.jcp.2011.04.019}.
  \DOIprefix\doi{10.1016/j.jcp.2011.04.019}.
\bibitem[{Kolobov and Arslanbekov(2012)}]{Kolobov_2012}
\bibinfo{author}{V.~Kolobov}, \bibinfo{author}{R.~Arslanbekov},
  \bibinfo{journal}{Journal of Computational Physics} \bibinfo{volume}{231}
  (\bibinfo{year}{2012}) \bibinfo{pages}{839--869}. \URLprefix
  \url{http://dx.doi.org/10.1016/j.jcp.2011.05.036}.
  \DOIprefix\doi{10.1016/j.jcp.2011.05.036}.
\bibitem[{Popinet(2003)}]{Popinet_2003}
\bibinfo{author}{S.~Popinet}, \bibinfo{journal}{Journal of Computational
  Physics} \bibinfo{volume}{190} (\bibinfo{year}{2003})
  \bibinfo{pages}{572--600}. \URLprefix
  \url{http://dx.doi.org/10.1016/S0021-9991(03)00298-5}.
  \DOIprefix\doi{10.1016/s0021-9991(03)00298-5}.
\bibitem[{Nijdam et~al.(2016)Nijdam, Teunissen, Takahashi, and
  Ebert}]{Nijdam_Teunissen_2016}
\bibinfo{author}{S.~Nijdam}, \bibinfo{author}{J.~Teunissen},
  \bibinfo{author}{E.~Takahashi}, \bibinfo{author}{U.~Ebert},
  \bibinfo{journal}{Plasma Sources Science and Technology} \bibinfo{volume}{25}
  (\bibinfo{year}{2016}) \bibinfo{pages}{044001}. \URLprefix
  \url{http://dx.doi.org/10.1088/0963-0252/25/4/044001}.
  \DOIprefix\doi{10.1088/0963-0252/25/4/044001}.
\bibitem[{Teunissen and Ebert(2017)}]{teunissen_2017_afivo_streamer}
\bibinfo{author}{J.~Teunissen}, \bibinfo{author}{U.~Ebert},
  \bibinfo{journal}{Journal of Physics D: Applied Physics} \bibinfo{volume}{50}
  (\bibinfo{year}{2017}) \bibinfo{pages}{474001}. \URLprefix
  \url{http://dx.doi.org/10.1088/1361-6463/aa8faf}.
  \DOIprefix\doi{10.1088/1361-6463/aa8faf}.
\bibitem[{Zhang et~al.(2016)Zhang, Almgren, Day, Nguyen, Shalf, and
  Unat}]{Zhang_2016}
\bibinfo{author}{W.~Zhang}, \bibinfo{author}{A.~Almgren},
  \bibinfo{author}{M.~Day}, \bibinfo{author}{T.~Nguyen},
  \bibinfo{author}{J.~Shalf}, \bibinfo{author}{D.~Unat}, \bibinfo{journal}{SIAM
  Journal on Scientific Computing} \bibinfo{volume}{38} (\bibinfo{year}{2016})
  \bibinfo{pages}{S156–S172}. \URLprefix
  \url{http://dx.doi.org/10.1137/15M102616X}.
  \DOIprefix\doi{10.1137/15m102616x}.
\bibitem[{Colella et~al.(2011)Colella, Graves, Johnson, Keen, Ligocki, Martin,
  McCorquodale, Modiano, Schwartz, Sternberg, and Straalen}]{chombo}
\bibinfo{author}{P.~Colella}, \bibinfo{author}{D.~T. Graves},
  \bibinfo{author}{J.~N. Johnson}, \bibinfo{author}{N.~D. Keen},
  \bibinfo{author}{T.~J. Ligocki}, \bibinfo{author}{D.~F. Martin},
  \bibinfo{author}{P.~W. McCorquodale}, \bibinfo{author}{D.~Modiano},
  \bibinfo{author}{P.~O. Schwartz}, \bibinfo{author}{T.~D. Sternberg},
  \bibinfo{author}{B.~V. Straalen}, \bibinfo{title}{Chombo software package for
  {AMR} applications - design document}, \bibinfo{year}{2011}. \URLprefix
  \url{https://apdec.org/designdocuments/ChomboDoc/ChomboDesign/chomboDesign.pdf}.
\bibitem[{Hornung et~al.(2006)Hornung, Wissink, and Kohn}]{Hornung_2006}
\bibinfo{author}{R.~D. Hornung}, \bibinfo{author}{A.~M. Wissink},
  \bibinfo{author}{S.~R. Kohn}, \bibinfo{journal}{Engineering with Computers}
  \bibinfo{volume}{22} (\bibinfo{year}{2006}) \bibinfo{pages}{181--195}.
  \URLprefix \url{http://dx.doi.org/10.1007/s00366-006-0038-6}.
  \DOIprefix\doi{10.1007/s00366-006-0038-6}.
\bibitem[{MacNeice et~al.(2000)MacNeice, Olson, Mobarry, de~Fainchtein, and
  Packer}]{Macneice_2000}
\bibinfo{author}{P.~MacNeice}, \bibinfo{author}{K.~M. Olson},
  \bibinfo{author}{C.~Mobarry}, \bibinfo{author}{R.~de~Fainchtein},
  \bibinfo{author}{C.~Packer}, \bibinfo{journal}{Computer Physics
  Communications} \bibinfo{volume}{126} (\bibinfo{year}{2000})
  \bibinfo{pages}{330--354}. \URLprefix
  \url{http://dx.doi.org/10.1016/S0010-4655(99)00501-9}.
  \DOIprefix\doi{10.1016/s0010-4655(99)00501-9}.
\bibitem[{Sampath et~al.(2008)Sampath, Adavani, Sundar, Lashuk, and
  Biros}]{Sampath_2008}
\bibinfo{author}{R.~S. Sampath}, \bibinfo{author}{S.~S. Adavani},
  \bibinfo{author}{H.~Sundar}, \bibinfo{author}{I.~Lashuk},
  \bibinfo{author}{G.~Biros}, \bibinfo{journal}{2008 SC - International
  Conference for High Performance Computing, Networking, Storage and Analysis}
  (\bibinfo{year}{2008}). \URLprefix
  \url{http://dx.doi.org/10.1109/SC.2008.5218558}.
  \DOIprefix\doi{10.1109/sc.2008.5218558}.
\bibitem[{Weinzierl(2009)}]{peano_book}
\bibinfo{author}{T.~Weinzierl}, \bibinfo{title}{A Framework for Parallel PDE
  Solvers on Multiscale Adaptive Cartesian Grids},
  \bibinfo{publisher}{Technische Universit{\"a}t M{\"u}nchen},
  \bibinfo{year}{2009}.
\bibitem[{Teyssier(2002)}]{Teyssier_2002}
\bibinfo{author}{R.~Teyssier}, \bibinfo{journal}{A\&A} \bibinfo{volume}{385}
  (\bibinfo{year}{2002}) \bibinfo{pages}{337--364}. \URLprefix
  \url{http://dx.doi.org/10.1051/0004-6361:20011817}.
  \DOIprefix\doi{10.1051/0004-6361:20011817}.
\bibitem[{Calhoun(2015)}]{www_donna_calhoun}
\bibinfo{author}{D.~Calhoun}, \bibinfo{title}{Adaptive mesh refinement
  resources}, \bibinfo{year}{2015}. \URLprefix
  \url{http://math.boisestate.edu/~calhoun/www_personal/research/amr_software/index.html},
  \bibinfo{note}{[Online; accessed 18-April-2018]}.
\bibitem[{Morton(1966)}]{Morton_1966}
\bibinfo{author}{G.~Morton}, \bibinfo{journal}{IBM Research Report}
  (\bibinfo{year}{1966}).
\bibitem[{et~al.(2015)}]{www_boxlib}
\bibinfo{author}{A.~S.~A. et~al.}, \bibinfo{title}{Boxlib},
  \bibinfo{year}{2015}. \URLprefix
  \url{https://ccse.lbl.gov/BoxLib/index.html}, \bibinfo{note}{[Online;
  accessed 22-July-2015]}.
\bibitem[{Carrard et~al.(2012)Carrard, Law, and P{\'e}bay}]{Carrard_2012}
\bibinfo{author}{T.~Carrard}, \bibinfo{author}{C.~Law},
  \bibinfo{author}{P.~P{\'e}bay}, \bibinfo{journal}{21st International Meshing
  Roundtable}  (\bibinfo{year}{2012}).
\bibitem[{Kitware(2015)}]{www_paraview}
\bibinfo{author}{Kitware}, \bibinfo{title}{Paraview}, \bibinfo{year}{2015}.
  \URLprefix \url{https://www.paraview.org/}, \bibinfo{note}{[Online; accessed
  20-April-2018]}.
\bibitem[{Childs et~al.(2012)Childs, Brugger, Whitlock, Meredith, Ahern,
  Pugmire, Biagas, Miller, Harrison, Weber, Krishnan, Fogal, Sanderson, Garth,
  Bethel, Camp, R\"{u}bel, Durant, Favre, and Navr\'{a}til}]{HPV:VisIt}
\bibinfo{author}{H.~Childs}, \bibinfo{author}{E.~Brugger},
  \bibinfo{author}{B.~Whitlock}, \bibinfo{author}{J.~Meredith},
  \bibinfo{author}{S.~Ahern}, \bibinfo{author}{D.~Pugmire},
  \bibinfo{author}{K.~Biagas}, \bibinfo{author}{M.~Miller},
  \bibinfo{author}{C.~Harrison}, \bibinfo{author}{G.~H. Weber},
  \bibinfo{author}{H.~Krishnan}, \bibinfo{author}{T.~Fogal},
  \bibinfo{author}{A.~Sanderson}, \bibinfo{author}{C.~Garth},
  \bibinfo{author}{E.~W. Bethel}, \bibinfo{author}{D.~Camp},
  \bibinfo{author}{O.~R\"{u}bel}, \bibinfo{author}{M.~Durant},
  \bibinfo{author}{J.~M. Favre}, \bibinfo{author}{P.~Navr\'{a}til}, in:
  \bibinfo{booktitle}{{High Performance Visualization--Enabling Extreme-Scale
  Scientific Insight}}, \bibinfo{publisher}{Chapman and Hall/CRC},
  \bibinfo{year}{2012}, pp. \bibinfo{pages}{357--372}.
\bibitem[{Brandt and Livne(2011)}]{Brandt_2011}
\bibinfo{author}{A.~Brandt}, \bibinfo{author}{O.~E. Livne},
  \bibinfo{title}{Multigrid Techniques}, \bibinfo{publisher}{Society for
  Industrial \& Applied Mathematics (SIAM)}, \bibinfo{year}{2011}. \URLprefix
  \url{http://dx.doi.org/10.1137/1.9781611970753}.
  \DOIprefix\doi{10.1137/1.9781611970753}.
\bibitem[{Trottenberg et~al.(2000)Trottenberg, Oosterlee, and
  Schuller}]{Trottenberg_2000_multigrid}
\bibinfo{author}{U.~Trottenberg}, \bibinfo{author}{C.~Oosterlee},
  \bibinfo{author}{A.~Schuller}, \bibinfo{title}{Multigrid},
  \bibinfo{publisher}{Elsevier Science}, \bibinfo{year}{2000}.
\bibitem[{Briggs et~al.(2000)Briggs, Henson, and McCormick}]{Briggs_2000}
\bibinfo{author}{W.~L. Briggs}, \bibinfo{author}{V.~E. Henson},
  \bibinfo{author}{S.~F. McCormick}, \bibinfo{title}{A Multigrid Tutorial
  (2\textsuperscript{nd} Ed.)}, \bibinfo{publisher}{Society for Industrial \&
  Applied Mathematics}, \bibinfo{address}{Philadelphia, PA, USA},
  \bibinfo{year}{2000}.
\bibitem[{Hackbusch(1985)}]{Hackbusch_1985}
\bibinfo{author}{W.~Hackbusch}, \bibinfo{journal}{Springer Series in
  Computational Mathematics}  (\bibinfo{year}{1985}). \URLprefix
  \url{http://dx.doi.org/10.1007/978-3-662-02427-0}.
  \DOIprefix\doi{10.1007/978-3-662-02427-0}.
\bibitem[{Bai and Brandt(1987)}]{Bai_1987}
\bibinfo{author}{D.~Bai}, \bibinfo{author}{A.~Brandt}, \bibinfo{journal}{SIAM
  Journal on Scientific and Statistical Computing} \bibinfo{volume}{8}
  (\bibinfo{year}{1987}) \bibinfo{pages}{109--134}. \URLprefix
  \url{http://dx.doi.org/10.1137/0908025}. \DOIprefix\doi{10.1137/0908025}.
\bibitem[{Arrayás and Ebert(2004)}]{Arrays_2004}
\bibinfo{author}{M.~Arrayás}, \bibinfo{author}{U.~Ebert},
  \bibinfo{journal}{Physical Review E} \bibinfo{volume}{69}
  (\bibinfo{year}{2004}). \URLprefix
  \url{http://dx.doi.org/10.1103/PhysRevE.69.036214}.
  \DOIprefix\doi{10.1103/physreve.69.036214}.
\bibitem[{Derks et~al.(2008)Derks, Ebert, and Meulenbroek}]{Derks_2008}
\bibinfo{author}{G.~Derks}, \bibinfo{author}{U.~Ebert},
  \bibinfo{author}{B.~Meulenbroek}, \bibinfo{journal}{J Nonlinear Sci}
  \bibinfo{volume}{18} (\bibinfo{year}{2008}) \bibinfo{pages}{551--590}.
  \URLprefix \url{http://dx.doi.org/10.1007/s00332-008-9023-0}.
  \DOIprefix\doi{10.1007/s00332-008-9023-0}.
\bibitem[{Ebert et~al.(2010)Ebert, Brau, Derks, Hundsdorfer, Kao, Li, Luque,
  Meulenbroek, Nijdam, Ratushnaya, and et~al.}]{Ebert_nonlin_2010}
\bibinfo{author}{U.~Ebert}, \bibinfo{author}{F.~Brau},
  \bibinfo{author}{G.~Derks}, \bibinfo{author}{W.~Hundsdorfer},
  \bibinfo{author}{C.-Y. Kao}, \bibinfo{author}{C.~Li},
  \bibinfo{author}{A.~Luque}, \bibinfo{author}{B.~Meulenbroek},
  \bibinfo{author}{S.~Nijdam}, \bibinfo{author}{V.~Ratushnaya},
  \bibinfo{author}{et~al.}, \bibinfo{journal}{Nonlinearity}
  \bibinfo{volume}{24} (\bibinfo{year}{2010}) \bibinfo{pages}{C1--C26}.
  \URLprefix \url{http://dx.doi.org/10.1088/0951-7715/24/1/C01}.
  \DOIprefix\doi{10.1088/0951-7715/24/1/c01}.
\bibitem[{Hagelaar and Pitchford(2005)}]{Hagelaar_2005}
\bibinfo{author}{G.~J.~M. Hagelaar}, \bibinfo{author}{L.~C. Pitchford},
  \bibinfo{journal}{Plasma Sources Science and Technology} \bibinfo{volume}{14}
  (\bibinfo{year}{2005}) \bibinfo{pages}{722--733}. \URLprefix
  \url{http://dx.doi.org/10.1088/0963-0252/14/4/011}.
  \DOIprefix\doi{10.1088/0963-0252/14/4/011}.
\bibitem[{Dujko et~al.(2011)Dujko, Ebert, White, and Petrovi\'{c}}]{Dujko_2011}
\bibinfo{author}{S.~Dujko}, \bibinfo{author}{U.~Ebert}, \bibinfo{author}{R.~D.
  White}, \bibinfo{author}{Z.~L. Petrovi\'{c}}, \bibinfo{journal}{Japanese
  Journal of Applied Physics} \bibinfo{volume}{50} (\bibinfo{year}{2011})
  \bibinfo{pages}{08JC01}. \URLprefix
  \url{http://dx.doi.org/10.1143/JJAP.50.08JC01}.
  \DOIprefix\doi{10.1143/jjap.50.08jc01}.
\bibitem[{Li et~al.(2010)Li, Ebert, and Hundsdorfer}]{Li_hybrid_i_2010}
\bibinfo{author}{C.~Li}, \bibinfo{author}{U.~Ebert},
  \bibinfo{author}{W.~Hundsdorfer}, \bibinfo{journal}{Journal of Computational
  Physics} \bibinfo{volume}{229} (\bibinfo{year}{2010})
  \bibinfo{pages}{200--220}. \URLprefix
  \url{http://dx.doi.org/10.1016/j.jcp.2009.09.027}.
  \DOIprefix\doi{10.1016/j.jcp.2009.09.027}.
\bibitem[{Rabie and Franck(2016)}]{Rabie_2016}
\bibinfo{author}{M.~Rabie}, \bibinfo{author}{C.~Franck},
  \bibinfo{journal}{Computer Physics Communications} \bibinfo{volume}{203}
  (\bibinfo{year}{2016}) \bibinfo{pages}{268–277}. \URLprefix
  \url{http://dx.doi.org/10.1016/j.cpc.2016.02.022}.
  \DOIprefix\doi{10.1016/j.cpc.2016.02.022}.
\bibitem[{Hockney and Eastwood(1988)}]{Hockney_1988}
\bibinfo{author}{R.~W. Hockney}, \bibinfo{author}{J.~W. Eastwood},
  \bibinfo{title}{{Computer simulation using particles.}},
  \bibinfo{publisher}{IOP Publishing Ltd.}, \bibinfo{address}{Bristol,
  England}, \bibinfo{year}{1988}.
\bibitem[{Dehnen and Read(2011)}]{Dehnen_2011}
\bibinfo{author}{W.~Dehnen}, \bibinfo{author}{J.~I. Read},
  \bibinfo{journal}{The European Physical Journal Plus} \bibinfo{volume}{126}
  (\bibinfo{year}{2011}). \URLprefix
  \url{http://dx.doi.org/10.1140/epjp/i2011-11055-3}.
  \DOIprefix\doi{10.1140/epjp/i2011-11055-3}.
\bibitem[{Ripperda et~al.(2018)Ripperda, Bacchini, Teunissen, Xia, Porth,
  Sironi, Lapenta, and Keppens}]{Ripperda_2018}
\bibinfo{author}{B.~Ripperda}, \bibinfo{author}{F.~Bacchini},
  \bibinfo{author}{J.~Teunissen}, \bibinfo{author}{C.~Xia},
  \bibinfo{author}{O.~Porth}, \bibinfo{author}{L.~Sironi},
  \bibinfo{author}{G.~Lapenta}, \bibinfo{author}{R.~Keppens},
  \bibinfo{journal}{The Astrophysical Journal Supplement Series}
  \bibinfo{volume}{235} (\bibinfo{year}{2018}) \bibinfo{pages}{21}. \URLprefix
  \url{http://dx.doi.org/10.3847/1538-4365/aab114}.
  \DOIprefix\doi{10.3847/1538-4365/aab114}.

\end{thebibliography}

\end{document}